\documentclass[11pt]{scrartcl}

\setkomafont{disposition}{\normalfont\bfseries}

\usepackage{setspace}
\usepackage{xcolor}
\usepackage[colorlinks,linkcolor=Blue,citecolor=Blue,
	urlcolor=Blue]{hyperref}
\usepackage{amssymb,latexsym,bm,mathrsfs,dsfont}
\usepackage{amsmath,amsfonts,amsthm}
\usepackage{natbib}  
\usepackage{textcomp}
\usepackage{comment}

\usepackage{epigraph}
\usepackage{graphics,graphicx}
\graphicspath{{graphics/}}
\usepackage{amsbsy}
\usepackage{cleveref}
\usepackage{enumitem}
\usepackage{booktabs} 
\usepackage{multirow} 
\usepackage{dutchcal}
\usepackage{afterpage}
\usepackage[many]{tcolorbox}
\usepackage{lipsum}
\usepackage{microtype}

\usepackage[linecolor=red,backgroundcolor=red!25, 
	bordercolor=red,prependcaption,textsize=tiny]{todonotes}

\usepackage{stmaryrd}   

\usepackage{abstract}
	\renewcommand{\abstractname}{}    
	 
\usepackage[en-US]{datetime2}		
	\DTMlangsetup{showdayofmonth=false}

\usepackage{subfig}

\usepackage[margin=1.0in]{geometry}
\definecolor{Blue}{rgb}{0.1,0.2,0.5}
\definecolor{Magenta}{cmyk}{0.1,0.9,0.0,0.3}
\definecolor{cites}{HTML}{324b13}
\definecolor{links}{HTML}{1a663b}

\definecolor{ocre}{RGB}{243,102,25} 
\definecolor{lightgray}{RGB}{229,229,229} 
\definecolor{mediumgray}{RGB}{120,120,120} 
\definecolor{darkblue}{rgb}{0.0, 0.18, 0.39}
\definecolor{burgundy}{rgb}{0.5, 0.0, 0.13}
\definecolor{armygreen}{rgb}{0.29, 0.33, 0.13}

\usepackage{pgfplots}
\pgfplotsset{width = 7cm, compat=1.5}
\usepackage{tikz}
\usetikzlibrary{fit}
\usetikzlibrary{calc}
\usepackage{color}
\usepackage{array}
\usepackage{multirow}
\usepgflibrary{arrows}
\pgfplotsset{compat=1.6}
\pgfplotsset{soldot/.style={color=black,only marks,mark=*}} 
\pgfplotsset{holdot/.style={color=black,fill=white,only marks,mark=*}}
\pgfplotsset{every axis/.append style={
                axis x line=center,
                axis y line=center,
                axis line style={->},
                xlabel={$x$},
                ylabel={$y$},
                line width=1pt,} }

\usepackage[T1]{fontenc}

\usepackage[sc]{mathpazo}    

\theoremstyle{plain}
\newtheorem{theorem}{Theorem}
\newtheorem{proposition}{Proposition}
\newtheorem{lemma}{Lemma}
\newtheorem{corollary}{Corollary}

\newtheorem{definition}{Definition}
\newtheorem{assumption}{Assumption}

\theoremstyle{definition}

\theoremstyle{remark}
\newtheorem{example}{Example}

\newtheorem*{example-1-c}{Example 1 (continued)}
\newtheorem*{example-2-c}{Example 2 (continued)}

\theoremstyle{plain}

\theoremstyle{remark}

\newcommand{\R}{\mathbb{R}}

\newcommand{\E}{\mathbb{E}}

\newcommand{\zero}{\boldsymbol{0}}
\newcommand{\BR}{B\hspace{-.2ex}R}  
\newcommand{\sth}{\,\colon\,}  
\newcommand{\andeq}{\qquad \text{and} \qquad}
  
\newcommand{\interior}{\operatorname{int}}

\newcommand{\cali}[1]{\mathcal{#1}}
\newcommand{\scr}[1]{\mathscr{#1}}

\newcommand{\argmax}{\mathrm{argmax}}

\providecommand{\possessivecite}[1]
	{\citeauthor{#1}'s\nolinebreak[2] (\citeyear{#1})}

\newcommand{\BF}{\boldsymbol{F}}

\newcommand{\BY}{\boldsymbol{Y}}

\newcommand{\BZ}{\boldsymbol{Z}}
\newcommand{\Bz}{\boldsymbol{z}}

\newcommand{\htheta}{\hat\theta}

\newcommand{\btheta}{\boldsymbol\theta}

\newcommand{\BTheta}{\boldsymbol{\Theta}}

\newcommand{\Mscript}{\scriptscriptstyle\cali M}

\newcommand{\thetaM}{\theta^{\Mscript}}

\newcommand{\bthetaM}{\btheta^{\Mscript}}

\newcommand{\sthetaM}{\theta^{\Mscript *}}
\newcommand{\bsthetaM}{\btheta^{\Mscript *}}
\newcommand{\bcircthetaM}{\btheta^{\Mscript \circ}}

\newcommand{\ThetaM}{\Theta^{\Mscript}}
\newcommand{\BThetaM}{\BTheta^{\Mscript}}

\newcommand{\Jscript}{\scriptscriptstyle\cali J}

\newcommand{\thetaJ}{\theta^{\Jscript}}

\newcommand{\hthetaJ}{\htheta^{\Jscript}}

\newcommand{\bthetaJ}{\btheta^{\Jscript}}

\newcommand{\sthetaJ}{\theta^{\Jscript *}}
\newcommand{\bsthetaJ}{\btheta^{\Jscript *}}

\newcommand{\ThetaJ}{\Theta^{\Jscript}}
\newcommand{\BThetaJ}{\BTheta^{\Jscript}}

\newcommand{\Escript}{\scriptscriptstyle\cali E}

\newcommand{\thetaE}{\theta^{\Escript}}

\newcommand{\bthetaE}{\btheta^{\Escript}}

\newcommand{\bsthetaE}{\btheta^{\Escript *}}

\newcommand{\ThetaE}{\Theta^{\Escript}}
\newcommand{\BThetaE}{\BTheta^{\Escript}}

\newcommand{\Zsat}{Z^{\mathrm{sat}}}
\newcommand{\zsat}{z^{\mathrm{sat}}}
\newcommand{\Bzsat}{\Bz^{\mathrm{sat}}}

\newcommand{\Bsig}{\boldsymbol{\sigma}}

\newcounter{BoxedDef}

\newcounter{BoxedProp}

\newcounter{BoxedThm}

\newcounter{BoxedLemma}

\newcounter{BoxedCor}

\newcounter{BoxedExample}

\begin{document}

\title{The Market and the Jungle%
	\thanks{Part of this work was written while the first 
	author visited \emph{Instituto de Estudios Peruanos} 
	in Lima -- Per\'u; 
	he is very thankful to his gracious hosts. 
	Financial support from the Australian Research Council 
	and the Australian Institute of Human Rights is 
	gratefully acknowledged. } } 

\author{
	Juan Carlos Carbajal%
		\footnote{School of Economics, UNSW Sydney, Australia.
		Email: \href{mailto:jc.carbajal@unsw.edu.au}
		{\nolinkurl{jc.carbajal@unsw.edu.au}}. }
	\and
	Marcelo Gallardo%
	\footnote{Department of Economics, UC Berkeley.
	Email: \href{mailto:marcelogallardob21@berkeley.edu}
	{\nolinkurl{marcelogallardob21@berkeley.edu}}. }
	\and
	Cesar Martinelli%
	\footnote{Interdisciplinary Center for Economic Science
		and Center for Public Choice, George Mason University.
	Email: \href{mailto:cmarti33@gmu.edu}
	{\nolinkurl{cmarti33@gmu.edu}}. }
}

\date{Preliminary Version -- \today}

\maketitle

\renewcommand{\abstractname}{}
\begin{abstract}

\noindent \hrulefill

\noindent {\bfseries Abstract \ }
We study how different institutions governing property rights 
and economic transactions shape occupational choices and 
welfare. 
Individuals in our society choose occupations that determine 
their production and consumption opportunities, 
anticipating either anonymous, voluntary market exchange 
supported by public enforcement (\emph{the market}), or 
allocations through a random coercive hierarchy 
(\emph{the jungle}) without monetary compensation.  
We establish conditions for occupational equilibria 
to exist in the market and in the jungle, and compare their 
performance.  
At a common occupational profile, the market maximizes 
aggregate surplus, whereas allocations in the jungle 
are Pareto efficient with monetary holdings fixed but 
need not maximize surplus.  
Endogenous occupational choices introduce an additional 
source of institutional differences.  
We find that a market occupational equilibrium outperforms 
a jungle occupational equilibrium, in aggregate terms, 
under two conditions: (i) every occupation in the jungle is 
available in the market, and (ii) the market is perfectly 
competitive, in the sense that prices cannot be manipulated 
by unilateral changes in occupations.  
We also establish an individual welfare comparison without 
this price-invariance condition: among identical individuals, 
everyone weakly prefers an occupational equilibrium in the market 
to a symmetric occupational equilibrium in the jungle. 
Our analysis places the institutional foundations of 
exchange at the center of the welfare comparison and provides 
a framework for studying coercive institutions that survive 
alongside markets.


\noindent \hrulefill

\end{abstract}


\section{Introduction}   \label{S:intro}

In this paper, we compare the allocation that obtains from 
a market economy ---for short, \emph{the market}--- with 
the allocation that obtains from an alternative mechanism 
based on the use of coercion ---for short, \emph{the jungle}.  
In the market, an allocation results from voluntary 
transactions among individuals based on prices and 
supported by public enforcement of property rights.  
In the jungle, an allocation results from the exercise of 
coercion in a hierarchy of power similar to that of a 
serial dictatorship.   
Individuals choose occupations in anticipation of the 
different rules governing each economy.  
Our concern is how these institutions shape both what 
individuals choose to do and the welfare they attain.

Individuals in our society have preferences that are 
quasilinear in a numeraire (money).  
A key feature of our model is that each individual is 
characterized by the set of occupations she has access to 
in each economy.  
These occupational opportunities, moreover, may or may not 
be the same.    
Occupations determine what individuals  
can do as consumers and as producers, as in 
\citet{Makowski:1995cf}: an occupation specifies a 
valuation function and a set of feasible trades. 
In the market, an allocation is reached concurrently with 
commodity prices via anonymous, decentralized trading and 
enforceable monetary transfers. 
In the jungle, more powerful individuals can decide the trading 
behavior of less powerful ones, as in \citet{Piccione:2007nh}.  
Thus, these institutions differ in both the way transactions 
are enforced and the scope for compensating individuals 
for their exchange activities.

Individuals choose their occupations before trading takes  
place in the market and before the power hierarchy is 
realized in the jungle. 
The hierarchy is drawn uniformly from all possible power 
orderings, independently of occupational choices.  
This captures uncertainty about the exercise of coercive 
power; individuals do not invest in their positions in 
the hierarchy, and their trading opportunities are not 
related to their coercive capacities.  
An occupational equilibrium is a profile at which no 
individual gains by changing her occupation, anticipating 
the resulting market payoff or expected jungle payoff.

We derive two main results.  

First, we consider a scenario in which the profile of 
occupations is the same in the market and in the jungle. 
We show that under those circumstances, individuals are 
collectively made weakly better off in the market than 
in the jungle, and under weak assumptions, strictly better off.  
This result rests upon well-known intuitions behind the 
properties of competitive equilibrium.  
At the fixed occupational profile, both market equilibrium 
allocation and jungle equilibrium allocation are locally 
Pareto efficient, but with quasilinear preferences and 
monetary transfers, only the market equilibrium maximizes 
aggregate payoffs, understood as the sum of valuations.  
Because the jungle lacks a public institution governing 
property rights and enforcing contracts, money supply is not 
available to compensate individuals and induce a 
surplus-enhancing allocation.

Second, allowing individuals to freely choose their occupations 
before transactions in the market and before power uncertainty 
resolves in the jungle, so that the equilibrium profile 
of occupations in both economies may differ, we show that 
individuals are collectively made better off in the market 
than in the jungle under two conditions: (i) every potential 
occupation in the jungle is also available in the market, 
and (ii) the market is perfectly competitive, in the sense 
that prices cannot be manipulated by any single individual, 
neither via trading decisions nor via occupational 
decisions.
We show that, when these two conditions are present, any 
occupational equilibrium in the market collectively 
outperforms any occupational equilibrium in the jungle. 
This aggregate comparison leaves open whether particular 
individuals prefer the jungle.  
We obtain an individual comparison for a special case, 
without requiring perfect competition in the occupational 
equilibrium in the market: among identical individuals, 
everyone weakly prefers an occupational equilibrium in 
the market to a symmetric occupational equilibrium in 
the jungle.   
This should be of independent interest.

Our first result parallels traditional welfare theorems in 
an economy with transferable utility.  
Our second result goes beyond, and provides a template to 
understand situations in which groups of individuals may opt 
out of the market. 
In particular, we interpret failures of condition (i) as 
circumstances in which due to social exclusion, stigma, 
or force, some occupations are only available in the jungle. 
We interpret failures of condition (ii) as corresponding to 
oligopolistic or monopolistic market structures in the market 
economy, in a broader sense than the standard interpretation 
in industrial organization models.  
In particular, the jungle economy may be preferable to the 
market economy for a group of individuals if certain occupations 
are available for them in the jungle but not in the market, 
or if prices can be rigged by consumption and/or production 
decisions to not fully reflect competitive forces, 
or if some individuals have occupations available to them 
that give them means to influence market prices.  
Thus, our model provides a flexible framework to explain, 
in terms of general equilibrium, the persistence of 
non-anonymous trading under duress and private enforcement, 
in spite of the superiority of the market mechanism.

Competitive markets in economic models are both an idealized 
representation of the complex process of allocating resources 
in a market economy with well-defined property rights 
and, via the welfare theorems, a representation of the 
ideal solution of the resource allocation problem. 
The state is often contraposed to the market as either, 
positively, a source of security in property rights, impartial 
enforcement of contracts, and a solution to potential 
market failures, or, negatively, as a source of deviations 
from the ideal market allocation due to unjustified 
government intervention. 
This to some extent betrays the origins of the discipline 
in the preoccupation with fiscal affairs and public policy 
of the classical political economists. 
In line with \citet{Basu:2003}, we think instead of the 
economy as deeply embedded in social institutions that 
constrain the full gamut of individual behaviors ---including 
occupational and trading decisions---  and the working 
of prices. 
These social institutions operate via formal state power 
or via social norms, traditions and the use of force. 
Indeed, examples of social exclusion due to ethnic, religious, 
or language differences are ubiquitous, though their 
importance in different economies can vary hugely. 
Exclusion, seizure of output and compelled labor whether or 
not mediated by the state are examples of what we 
interpret as primary conditions for the persistence of an 
alternative economic system opposing the market, which we 
call the jungle.

We eschew the traditional discussion of market shortcomings 
as ``market failures.''  
In particular, we do not counterpose the market to the 
idealized intervention of a benevolent state that 
corrects market shortcomings. 
Rather, we are interested in comparing allocations 
predicted by the market mechanism with those dictated by 
the alternative allocation by coercive force, and ask 
under what circumstances the second may afford some 
individuals a better prospect ex ante.  
Indeed, coercive power can itself sustain economic order when 
those who exercise it have an interest in the resources 
generated by production and exchange \citep{Olson:1993ni}.%
		\footnote{\citet{SanchezDeLaSierra:2020rc} documents 
		how armed actors in eastern Congo establish systems 
		of taxation and protection, and finds that some forms 
		of armed rule improve household welfare.} 
Our framework allows us to pin down, in the language of general 
equilibrium expanded to include occupational choices, precise 
conditions for the coexistence of the market and the jungle.  
Thus, our paper highlights not just the importance of 
price-taking behavior, including in occupational choices, but 
also the role of market exclusion via social norms, traditions, 
and state regulations.

The remainder of the paper is organized as follows.  
\autoref{S:framework} describes the model of our society, 
one in which individuals can choose whether to join the 
market or the jungle and what occupation to select. 
\autoref{S:jungle} describes in detail the workings of the jungle 
economy, and provides conditions for the existence of an 
occupational equilibrium in the jungle.  
Turning to the market economy, \autoref{S:market} provides 
conditions for the existence of an occupational equilibrium 
in the market.   
At any occupational profile, a local market equilibrium 
exists and is Pareto efficient. 
Likewise, at any occupational profile, a serial dictatorship 
procedure selects local jungle equilibria that are Pareto 
efficient when monetary holdings are fixed.  
Passing from these allocations to occupational equilibrium
requires additional care.  
We illustrate some of these difficulties with a few examples. 
\autoref{S:comparison} compares the outcomes of both 
economies and provides our two main results described 
above.


\section{Setting}   \label{S:framework}

We start with a general description of a society as a 
finite group of individuals who make occupational choices 
and engage in production and consumption activities. 
Some of these individuals interact in the \emph{market}, 
some of them interact in the \emph{jungle}. 
Individuals have a choice as to which economy they belong to.
In this paper, to simplify the analysis we take the initial 
participation choices to be irrevocable.%
		\footnote{Our focus on entry decisions highlights 
		individual agency, but entry and exit also face 
		frictions. 
		\citet{melnikov2025gangs} show that criminal organizations 
		in El Salvador restrict individuals' mobility and 
		labor-market options; \citet{sviatschi2022narco} 
		documents the persistence of criminal career paths.}
These economies operate under different rules, which
we explain below. 
The crucial distinction between them lies in their enforcement
of economic transactions: anonymous exchange in the market 
relies on public enforcement of property rights and contracts; 
the coordination of economic activity in the jungle relies 
instead on private enforcement by a hierarchy.

The set of individuals who belong to the market is $\cali M$,
the set of individuals who belong to the jungle is $\cali J$.
We assume $\cali M$ and $\cali J$ to be disjoint and let
$\cali S = \cali M \cup \cali J$ denote the set of all
individuals in our society.
In explaining our framework, we treat market and jungle
as nonempty.
With endogenous participation, it is possible that
$\cali M = \emptyset$ or that $\cali J = \emptyset$
(but both cannot be empty).
We are particularly interested in the case where these two
economies coexist, i.e., $\cali M \ne \emptyset$ and
$\cali J \ne \emptyset$ simultaneously.

We assume that individuals have convex, quasilinear preferences 
over $L$ non-monetary commodities and money. 
Think of the $L$ commodities as indexing all the goods 
and services that could be offered in this society.  
Some goods may be actively traded mostly in the market, 
others mostly in the jungle, and some in both economies.  
All individuals in this society act both as consumers 
and entrepreneurs.
To capture this in a flexible way, we take a cue from
\citet{Makowski:1995cf} and let $z_i \in \R^L$ be a 
\emph{trade} for each $i \in \cali S$, with 
the sign convention that treats $z_{i\ell} > 0$ as 
the amount of good $\ell$ being purchased and $z_{i\ell} < 0$ 
as the amount being sold by individual $i$. 
Inequalities between vectors are component-wise throughout.

Each individual $i\in\cali S$ is endowed with a nonempty set 
of \emph{occupations} $\Theta_i$.%
		\footnote{One interpretation: there are $K$ 
		productive activities available to $i$ in this society, 
		and $\theta_{ik}$ is the share of time that $i$ devotes 
		to activity $k$.
		So $\Theta_i$ is the simplex $\{\theta_i \in \R^{K}_+ 
		\sth \sum_{k=1}^K \theta_{ik} \le 1\}$. 
		Nothing below depends on this reading.} 
Occupations together with trades determine individual 
payoffs. 
A trade $z_i$ and a money holding $m_i\in\R$ give 
individual $i$ a utility of $v_i(z_i,\theta_i) \, + \, 
m_i$, where 
	\begin{equation*}
		v_i \colon \R^L \times \Theta_i \to \R 
		\cup \{-\infty\} .
	\end{equation*}
The cardinal normalization is part of the primitives and is 
fixed across occupations, so payoffs are comparable across 
the occupational choices available to an individual.  
Money is the unit of account in both economies, although, as we 
shall see later, only the market trades the numeraire.

An occupation $\theta_i \in \Theta_i$ also determines the 
set of individually feasible trades
	\begin{equation*}
		Z_i(\theta_i) \ \equiv \ 
		\{z_i \in \R^L \sth v_i(z_i,\theta_i) \, > \, -\infty\} .	
	\end{equation*}
$Z_i(\theta_i)$ records what individual $i$ can consume and 
produce at her given occupational choice.  
The value $-\infty$ outside $Z_i(\theta_i)$ encodes infeasibility. 
The family of induced valuation functions for the 
non-monetary commodities is $\{v_i(\cdot,\theta_i)
\sth\theta_i\in\Theta_i\}$.  
We consider the mapping $\theta_i \mapsto v_i(\cdot,\theta_i)$ 
to be injective, so the parameter $\theta_i$ and the induced 
valuation function identify the same occupation.

For the remainder of the paper, the following is maintained. 

\begin{assumption} \label{A:structure} 
For each $i \in \cali S$, assume that  
\begin{enumerate}[leftmargin=1.5em,label=(\alph*)] 
		\setcounter{enumi}{0} \itemsep.25ex
	\item the set of occupations $\Theta_i$ is a nonempty, 
		compact and convex subset of a finite-dimensional 
		normed vector space; 
	\item the correspondence $Z_i \colon \Theta_i \rightrightarrows 
		\R^L$ is continuous with nonempty, compact, and 
		convex values, and moreover 
		$\zero \in Z_i(\theta_i)$, for all $\theta_i \in 
		\Theta_i$; 
	\item the function $v_i \colon \R^L \times \Theta_i \to 
		\R \cup \{- \infty\}$ is continuous on $\big\{ 
		(z,\theta_i) \sth z \in Z_i(\theta_i), \, \theta_i 
		\in \Theta_i \big\}$. 
	\item Finally, $\zero \in \interior 
		\big\{ \sum_{i \in \cali S} Z_i(\theta_i) \big\}$,
		for all $\btheta = (\theta_i) \in \BTheta = 
		\times_{i \in \cali S} \Theta_i$.
\end{enumerate}
\end{assumption}

Most of the conditions stated in \autoref{A:structure} 
are standard in parametric optimization problems, and 
we will employ some of these tools in our analysis. 
We point out that, since individuals in our society have 
convex, quasilinear preferences over commodities and the 
numeraire, $v_i(\cdot,\theta_i)$ is concave on 
$Z_i(\theta_i)$, for all $i \in \cali S$. 
We explain the role our conditions play in due course.

In our model, individuals choose occupations in the market 
and in the jungle.  
The initial participation decision, whether to belong to 
the jungle or to the market, is influenced by the 
occupations an individual has available in each of these 
economies, which may differ due to formal regulations 
in the market, and customs or social norms in the market and 
in the jungle.   
To capture this, we let $\ThetaM_i \subseteq \Theta_i$ and 
$\ThetaJ_i \subseteq \Theta_i$ denote the (nonempty) sets 
of occupations available for $i \in \cali S$ if she chooses 
to join the market or the jungle, respectively.
In some contexts, it is possible that $\ThetaM_i = \ThetaJ_i 
= \Theta_i$.  
Other contexts may allow individual $i$ to attain some occupations
in the jungle that are not available for her in the market,
and vice versa.
Our framework is sufficiently flexible to capture these
possibilities.  
Let $\BThetaM = \times_{i \in \cali M} \ThetaM_i$ and
$\BThetaJ = \times_{i \in \cali J} \ThetaJ_i$.
An \emph{occupational profile} for the economy  $\cali E \in 
\left\{\cali M, \cali J\right\}$, denoted $\bthetaE = 
(\thetaE_i) \in \BThetaE$,  specifies an occupational 
choice for each individual in the given economy. 
We impose \autoref{A:structure} on each nonempty economy 
as well, replacing $\cali S$ and $\Theta_i$ by $\cali E$ and
$\ThetaE_i$, respectively.

A \emph{trade profile} $\Bz = (z_i)_{i \in \cali E}$ for the
economy $\cali E \in \left\{\cali M, \cali J\right\}$ is 
a collection of trades, one for each member of the 
economy.  
A first requirement for a trade profile to be feasible 
in $\cali E$ is that each trade is individually 
feasible given the occupational choices selected by 
each member of that economy.
A second requirement is that trades among all
individuals balance out.
We impose these two conditions throughout.

\begin{definition}   \label{D:feasible-trades}
Given an occupational profile $\bthetaE \in \BThetaE$ for the
economy $\cali E \in \left\{ \cali M, \cali J\right\}$, a trade
profile $\Bz = (z_i)_{i \in \cali E}$ is \emph{feasible
at $\bthetaE$} if the following two conditions are satisfied:
\begin{enumerate}[leftmargin=1.5em,label=(\alph*)]
		\setcounter{enumi}{0} \itemsep.25ex
	\item for all $i \in \cali E$, $z_i \in Z_i(\thetaE_i)$;
	\item $\sum_{i \in \cali E} z_i \, = \, \zero$.
\end{enumerate}
\end{definition}

The clearing condition in \autoref{D:feasible-trades}(b) 
is imposed within each economy, so the market and the 
jungle clear separately, with no trade between them.
Let $\BZ(\bthetaE)$ denote the set of all trade profiles for 
$\cali E$ that are feasible at $\bthetaE \in \BThetaE$.  
We state the next simple observation for completeness.

\begin{lemma}  \label{L:feasible-trades}
Let $\cali E \in \left\{\cali M, \cali J\right\}$ be a nonempty 
economy.
For any occupational profile $\bthetaE \in \BThetaE$, the set of 
feasible trade profiles $\BZ(\bthetaE) \subseteq 
\R^{|\cali E| \times L}$ is nonempty, compact and convex.
\end{lemma}

\begin{proof}
Let $\BF \, \equiv \, \big\{ \Bz = (z_i) \in \R^{|\cali E| 
\times L} \sth \sum_i z_i = \zero \big\}$ represent 
the trading feasibility constraint in $\cali E$. 
For any occupational profile $\bthetaE = (\thetaE_i) \in 
\BThetaE$, we can express the set of feasible trades as 
	\begin{align*}
		\BZ(\bthetaE) 
		\ = \ \Big[ \times_{i \in \cali E} Z_i(\thetaE_i) \Big] 
			\, \cap \, \BF .
	\end{align*}
Clearly, $\BZ(\bthetaE)$ is nonempty, either by 
\autoref{A:structure}(b) or \autoref{A:structure}(d). 
Further, by \autoref{A:structure}(b), $Z_i(\thetaE_i)$ is 
compact and convex for all $i \in \cali E$, and thus so is 
the product set $\times_{i \in \cali E} Z_i(\thetaE_i)$.  
Since the feasibility constraint $\BF$ is closed and convex, it 
follows that $\BZ(\bthetaE)$ is compact and convex.  
\end{proof}

To evaluate the performance of these economies, once 
occupational choices have been made, we employ a 
Pareto efficiency criterion with respect to the 
non-monetary commodities.

\begin{definition}  \label{D:local-efficiency}
Given an occupational profile $\bthetaE \in \BThetaE$ for the
economy $\cali E \in \left\{ \cali M, \cali J\right\}$, a trade
profile $\Bz = (z_i)_{i \in \cali E}$ is \emph{Pareto
efficient at $\bthetaE$} if it is feasible at $\bthetaE$,
i.e., $\Bz \in \BZ(\bthetaE)$, and if there is no other trade
profile $\hat \Bz = (\hat z_i)_{i \in \cali E}$, 
feasible at $\bthetaE$, such that $v_i (\hat z_i, \thetaE_i) 
\geq v_i (z_i, \thetaE_i)$ for all $i \in \cali E$ and $v_j 
(\hat z_j, \thetaE_j) > v_j (z_j, \thetaE_j)$ for at least some 
$j \in \cali E$.
\end{definition}

Our notion of efficiency evaluates trade profiles keeping 
monetary holdings fixed. 
In the jungle, this is the relevant criterion, since 
monetary transfers are excluded: there is no anonymous 
institution enforcing contracts and organizing  
monetary supply. 
In the market, in contrast, allowing zero-sum monetary transfers 
gives a stronger notion of efficiency: under quasilinear 
preferences, a feasible allocation of commodities and money is 
Pareto efficient if and only if its trade profile maximizes 
the sum of individual valuations.  
Such a trade profile is also Pareto efficient in the sense 
of \autoref{D:local-efficiency}, but the converse need 
not hold.  
We distinguish between these two notions in our analysis.


\section{The Jungle}  \label{S:jungle}

A main distinction between the jungle and the market is the 
way each economy enforces contracts. 
The market supports anonymous exchange and enforceable 
monetary transfers. 
In the jungle economy, there exists no third-party or public 
institution governing property rights. 
Thus, the production and consumption of resources in
the jungle are purely based on private enforcement without 
monetary compensations.

For this section, we take $\cali J \ne \emptyset$ to be given.
The coordination of economic activity in the jungle is 
organized by a hierarchy (or strict power relation).  
We model a hierarchy as a bijection
	\begin{equation*}
		h \colon \left\{1,2,\ldots,|\cali J|\right\} 
		\, \to \, \cali J 		
	\end{equation*}
that maps each position $n \in \{1,2,\ldots,|\cali J|\}$ 
in the hierarchy to the individual $h(n)$ who occupies it. 
Thus $h(1)$ is the most powerful individual, $h(2)$ the 
second most powerful, and so on. 
The interpretation is that, for any two positions $1 \leq n 
< m \leq |\cali J|$, individual $h(n)$ comes before $h(m)$ 
in the hierarchy $h$ and therein controls, coercively 
if needed, any available resource that $h(m)$ may have. 
Let $\scr H(\cali J)$ denote the set of all hierarchies 
on $\cali J$.

We assume that the exercise of coercive power is costless; 
for a discussion of this assumption, see \cite{Piccione:2007nh}. 
This is the counterpart of the frictionless exercise of property 
rights in a standard general equilibrium framework, 
which we will use to analyze the market economy.  
We record a power relation in the jungle as a bijection rather than 
as an abstract binary relation because it is the object over which 
we later randomize: a power relation is a permutation of $\cali J$, 
so a lottery over these hierarchies is a lottery over the
finite set $\scr H(\cali J)$.

\subsection{Local Equilibrium in the Jungle}
		\label{SS:local-jungle-equilibrium}

An equilibrium captures the idea of a stable social situation.  
In the jungle, consumption and production decisions are 
determined in a hierarchical way and enforced by private 
coercion.  
To capture these features of the workings of the jungle economy, 
we use a modified version of \possessivecite{Piccione:2007nh} 
jungle equilibrium.

\begin{definition}  \label{D:local-j-equilibrium}
Fix an occupational profile $\bthetaJ \in \BThetaJ$ and
a hierarchy $h \in \scr H (\cali J)$.
A trade profile $\Bz$ is a \emph{jungle equilibrium at 
$(\bthetaJ,h)$} if it satisfies
the following conditions:
\begin{enumerate}[leftmargin=1.5em,label=(\alph*)]
		\setcounter{enumi}{0} \itemsep.25ex
	\item $\Bz$ is feasible at $\bthetaJ$, i.e.,
		$\Bz \in \BZ(\bthetaJ)$;
	\item there do not exist two individuals $h(m),h(n)
		\in \cali J$, with $1 \leq m < n \leq |\cali J|$, 
		and a feasible trade profile
		$\hat\Bz \in \BZ (\bthetaJ)$, such that 	$v_i 
		(\hat z_i,\thetaJ_i) \, = \, v_i (z_i,\thetaJ_i)$  
		for all $i \in \cali J$, $i \ne 	h(m),h(n)$, and 
		further
		\vspace{-1ex}
		\begin{equation*}
			v_{h(m)} \big( \hat z_{h(m)}, \thetaJ_{h(m)} \big) 
			\ > \ v_{h(m)} \big( z_{h(m)}, \thetaJ_{h(m)} \big).
		\end{equation*}
\end{enumerate}
\end{definition}

We sometimes refer to a jungle equilibrium at a given occupational
profile and power relation as a \emph{local jungle equilibrium}.
Our definition of local jungle equilibria is stated in
terms of utilities instead of trades.
For an individual to ``break an equilibrium,'' she needs to
find an alternative feasible trade profile that makes her
strictly better off.
She can change trades across multiple individuals
in the jungle, but must do so in a way that leaves everyone
else, except a single weaker individual (in accordance with 
the hierarchy $h$), with the same utility level as in 
the original trade profile.  
Ours is stronger than the original one in \citet{Piccione:2007nh}, 
as their definition requires an individual $h(m)$ to change 
her trade and that of a single, weaker 
individual $h(n)$, leaving everyone else with the same trades.  
\autoref{D:local-j-equilibrium} fixes the other individuals 
only in \emph{utility} terms: the trades of third parties, 
including those above $h(m)$ in the hierarchy, may be 
rearranged along their indifference sets, but no third 
party's welfare changes.

\autoref{D:local-j-equilibrium} can be interpreted as a 
\emph{coercive blocking rule}, the power analogue of a 
blocking coalition: an individual $h(m)$ may costlessly 
propose any feasible reshuffle that strictly raises her 
own payoff and leaves every individual other than $h(m)$ 
and a single victim $h(n)$, ranked lower in the hierarchy 
(so $m < n$), at their original utility levels.
This is a reduced form of costless coercion in a moneyless 
economy.

We now show that a local jungle equilibrium exists.
Adapting the serial dictatorship procedure, we use the strict
order given by the hierarchy $h$ to define, recursively, 
a sequence of nested individual optimization problems.  
The critical property of these nested problems is that  
the solution set in one stage becomes the constraint set of 
the next stage.  
Since $|\cali J| < \infty$, the recursion is well defined and
produces a nonempty set of feasible trade profiles, each of
them constitutes a local jungle equilibrium.

\begin{proposition}  \label{P:local-j-existence}
For any occupational profile $\bthetaJ \in \BThetaJ$ and any
hierarchy $h \in \scr H (\cali J)$, a local jungle 
equilibrium exists.
\end{proposition}

\begin{proof}
Fix an occupational profile $\bthetaJ$ and a hierarchy $h$, 
and consider the following recursion. 

\smallskip
\noindent \emph{Initial Step.}  
Define $\BZ_{h(0)} (\bthetaJ) \equiv \BZ (\bthetaJ)$ as   
the constraint set of the most powerful individual.  
From \autoref{L:feasible-trades}, $\BZ_{h(0)}$ is nonempty 
and compact.  
Individual $h(1)$'s optimization problem is to choose a 
trade profile to maximize her utility among all feasible 
trade profiles, i.e., 
	\begin{equation*}
		\max_{\Bz} \left\{ v_{h(1)} \big(z_{h(1)},\thetaJ_{h(1)} 
			\big) \sth \Bz \, \in \, \BZ_{h(0)}(\bthetaJ) \right\} .
	\end{equation*}
Let $\pi_{h(1)} (\bthetaJ)$ denote the value of $h(1)$'s utility
at any solution of the above problem.
Define
	\begin{equation*}
		\BZ_{h(1)} (\bthetaJ)
		\ \equiv \ \left\{ \Bz \in \BZ_{h(0)} (\bthetaJ)
			\sth v_{h(1)} \big(z_{h(1)},\thetaJ_{h(1)}\big) 
			\, = \, \pi_{h(1)} (\bthetaJ) \right\} 
	\end{equation*}
as the solution set of $h(1)$'s optimization problem.  
Clearly, from \autoref{A:structure}(c), $\BZ_{h(1)} (\bthetaJ)$ 
is a nonempty, compact subset of $\BZ_{h(0)} (\bthetaJ)$.  

\smallskip
\noindent \emph{Induction Step.}
For $2 \leq n \leq |\cali J|$, consider the optimization
problem of individual $h(n) \in \cali J$.
She chooses a feasible trade profile to maximize her utility
among all feasible trade profiles that leave every stronger
individual $h(m)$, $m < n$, at a utility level equal to
$\pi_{h(m)} (\bthetaJ)$. 
Thus, $h(n)$'s optimization problem is
	\begin{equation*}
		\max_{\Bz} \left\{ v_{h(n)} \big(z_{h(n)},\thetaJ_{h(n)} 
			\big) \sth \Bz \, \in \, \BZ_{h(n-1)}(\bthetaJ) \right\} .
	\end{equation*}
Let $\pi_{h(n)}(\bthetaJ)$ denote the value of $h(n)$'s utility
at any solution of the above problem, and let
	\begin{equation*}
		\BZ_{h(n)} (\bthetaJ)
		\ \equiv \ \left\{ \Bz \in \BZ_{h(n-1)}(\bthetaJ)
			\sth v_{h(n)} \big(z_{h(n)},\thetaJ_{h(n)} \big) 
			\, = \, \pi_{h(n)} (\bthetaJ) \right\} .
	\end{equation*}
Clearly, the solution set $\BZ_{h(n)} (\bthetaJ)$ for 
individual $h(n)$'s optimization problem is a nonempty, 
compact subset of $\BZ_{h(n-1)} (\bthetaJ)$.  

\smallskip 
\noindent \emph{Terminal Step.}
Let $\BZ^*(\bthetaJ,h) \, \equiv \, \BZ_{h(|\cali J|)}(\bthetaJ)$.  
This set is nonempty and compact: it is the last of the 
finitely many solution sets constructed above, each of which 
is the set of maximizers of a continuous function over 
a nonempty compact set.

\smallskip

We claim that any trade profile $\Bz^* \in \BZ^* (\bthetaJ,h)$
is a jungle equilibrium at $(\bthetaJ,h)$.  
First, observe that the solution sets of the individual
optimization problems in the above recursion are nested:
	\begin{equation*}  
		\BZ^*(\bthetaJ,h) \ \equiv \
		\BZ_{h(|\cali J|)} (\bthetaJ)
		\ \subseteq \ \ldots \ \subseteq \
		\BZ_{h(1)} (\bthetaJ) \ \subseteq \ 
		\BZ_{h(0)} (\bthetaJ) \ \equiv \ \BZ(\bthetaJ). \
	\end{equation*}
Thus, $\Bz^* \in \BZ^*(\bthetaJ,h)$ is a feasible trade profile
for $\bthetaJ$.
To obtain a contradiction, suppose there is another
trade profile $\hat \Bz \in \BZ (\bthetaJ)$
and two individuals $h(m),h(n)$, with $m < n$, such that
$v_{i} (\hat z_{i},\thetaJ_{i}) \, = \, v_{i} (z^*_{i},
\thetaJ_{i})$ for all $i \in \cali J$, $i \ne h(m),h(n)$, 
and moreover 
	\begin{equation*}
		v_{h(m)} \big(\hat z_{h(m)},\thetaJ_{h(m)} \big) 
		\ > \ v_{h(m)} \big(z^*_{h(m)},\thetaJ_{h(m)} \big) .		
	\end{equation*}

Since $h(m)$ is more powerful than $h(n)$, the trade profile
$\hat \Bz$ must leave every stronger individual $h(k)$, for 
$k < m < n$, at their original utility level.
Hence the trade profile $\hat \Bz$ satisfies the constraints
defining each of the first $m-1$ optimization problems 
in our recursion.
It follows that $\hat \Bz \in \BZ_{h(m-1)} (\bthetaJ)$.
But this means $\hat \Bz$ is feasible in $h(m)$'s optimization
problem whose maximum value is $\pi_{h(m)} (\bthetaJ)$, so
we must have 
	\begin{equation*}
		v_{h(m)} \big(\hat z_{h(m)},\thetaJ_{h(m)} \big) 
		\ \le \ \pi_{h(m)}(\bthetaJ) 
		\ = \ v_{h(m)} \big(z^*_{h(m)},\thetaJ_{h(m)} \big) ,
	\end{equation*}
which is a contradiction.
\end{proof}

The construction in \autoref{P:local-j-existence} is 
an adaptation of the serial dictatorship procedure.
In house allocation, agents pick in a fixed order among the 
objects that remain; an assignment is Pareto efficient if 
and only if it is the outcome of some serial dictatorship, 
and uniform randomization over priority orders 
coincides in distribution with the core from uniformly 
random endowments \citep{AbdulkadirogluL1998tw}.
Axiomatizations and extensions are given by
\citet{Papai:2000jt} and \citet{Bade:2020ad}.
The closest antecedent in a continuous environment is
\citet{Satterthwaite:1981bk}.
Our serial dictators are constrained by their predecessors' 
\emph{utility levels}, not their assigned positions; under the 
strict preferences of the housing model the distinction 
is moot.
With endowments and consent the housing model
with strict preferences yields top
trading cycles \citep{Shapley:1974lh,Roth:1977wd,Roth:1982ic};
removing both and replacing them with the power order gives the
serial dictatorship above.  
\citet{Balbuzanov:2019bu} model rights to \emph{exclude} rather
than power to take; their exclusion core is a cooperative
counterpart of the blocking rule behind the local jungle 
equilibrium of \autoref{D:local-j-equilibrium}.

In proving \autoref{P:local-j-existence}, we identify a 
family $\BZ^*(\bthetaJ,h)$ of equilibrium trade profiles.  
Since individual preferences are convex, $\BZ^*(\bthetaJ,h)$ 
is a convex set,  within which the equilibria are 
utility-equivalent.

\begin{corollary}   \label{C:local-j-properties}
Fix an occupational profile $\bthetaJ \in \BThetaJ$ and a power 
relation $h \in \scr H(\cali J)$.
\begin{enumerate}[leftmargin=1.5em,label=(\alph*)]
		\setcounter{enumi}{0} \itemsep.25ex
	\item $\BZ^*(\bthetaJ,h)$ is convex.
	\item If $\Bz^*$ and $\Bz^{**}$ are two local jungle 
		equilibria in $\BZ^* (\bthetaJ,h)$, then for every 
		$i \in \cali J$, one has $v_{i} (z^*_{i},\thetaJ_{i}) 
		\, = \, v_{i} (z^{**}_{i},\thetaJ_i)$. 
	\item If $v_i(\cdot,\thetaJ_i)$ is strictly concave
		for all $i \in \cali J$, then $\BZ^*(\bthetaJ,h)$ is 
		a singleton. 
\end{enumerate}
\end{corollary}

\begin{proof}
(a) We show by induction on the stages $1 \leq n \leq 
|\cali J|$ that each solution set $\BZ_{h(n)}(\bthetaJ)$ 
is convex, starting from $\BZ_{h(0)}(\bthetaJ) = \BZ(\bthetaJ)$, 
which is convex by \autoref{L:feasible-trades}.  
Suppose now that $\BZ_{h(n-1)}(\bthetaJ)$ is convex.  
Let $\Bz',\Bz'' \in \BZ_{h(n)}(\bthetaJ)$, and let $\Bz^\lambda 
\, = \, \lambda\Bz' + (1-\lambda)\Bz''$, for any 
$\lambda\in[0,1]$.  
Then $\Bz^\lambda\in\BZ_{h(n-1)}(\bthetaJ)$, and using 
the concavity of $v_{h(n)} \big(\cdot,\thetaJ_{h(n)}\big)$ 
we obtain
	\begin{equation}  \label{Eq:local-j-properties}
	\begin{aligned}
		v_{h(n)} \big( z_{h(n)}^\lambda,\thetaJ_{h(n)} \big)
		& \ \ge \ \lambda v_{h(n)} \big(z_{h(n)}',
			\thetaJ_{h(n)} \big) \ + \ 
			(1-\lambda) v_{h(n)} \big(z_{h(n)}'', 
			\thetaJ_{h(n)} \big)   \\
		& \ = \ \pi_{h(n)}(\bthetaJ).
	\end{aligned}
	\end{equation}
Since $\pi_{h(n)}(\bthetaJ)$ is the value of individual $h(n)$'s 
maximized utility over $\BZ_{h(n-1)}(\bthetaJ)$, equality holds 
in the above expression and thus $\Bz^\lambda \in 
\BZ_{h(n)}(\bthetaJ)$.

\smallskip

(b) This is immediate from the recursion in the proof of 
\autoref{P:local-j-existence}.

\smallskip

(c) If any two trade profiles in $\BZ_{h(n)}(\bthetaJ)$ had 
different coordinates for individual $h(n)$, strict 
concavity of the valuation would make the inequality in 
\autoref{Eq:local-j-properties} strict for $0 < \lambda < 1$, 
which is impossible. 
Hence all trade profiles in $\BZ_{h(n)}(\bthetaJ)$ share 
the coordinate for $h(n)$. 
Since $\BZ_{h(n)}(\bthetaJ) \subseteq \BZ_{h(n-1)}(\bthetaJ)$, 
it follows that the coordinates of $h(1),\ldots,h(n-1)$ must 
also be the same, for any two trade profiles in 
$\BZ_{h(n)}(\bthetaJ)$.  
At the last stage of the recursion, every coordinate other 
than that of $h(|\cali J|)$ is fixed, and the valuation of 
the weakest individual is strictly concave. 
It follows that $\BZ^*(\bthetaJ,h)$ is a singleton.
\end{proof}

There may be additional local jungle equilibria outside
$\BZ^*(\bthetaJ,h)$, and these need not share the utility 
levels identified in \autoref{P:local-j-existence}. 
Therefore, the utility equivalence recorded in 
\autoref{C:local-j-properties}(b) is confined to the equilibrium 
class $\BZ^*(\bthetaJ,h)$.  
We also point out that the uniqueness identified in 
\autoref{C:local-j-properties}(c) concerns the family 
$\BZ^*(\bthetaJ,h)$.
Strict concavity alone does not exclude local jungle 
equilibria outside that family. 
On the other hand, as we show in our next result, 
the local jungle equilibria in $\BZ^*(\bthetaJ,h)$ are 
Pareto efficient at $\bthetaJ$.  
This conclusion follows from the serial selection
procedure; the local equilibrium definition alone does not
guarantee Pareto efficiency. 
At each stage in the recursion used to obtain 
$\BZ^*(\bthetaJ,h)$, an individual maximizes her utility 
among trade profiles preserving the utility levels obtained 
by individuals at the preceding stages.

\begin{proposition}   \label{P:local-j-efficiency}
Given an occupational profile $\bthetaJ \in \BThetaJ$ and 
a power relation $h \in \scr H(\cali J)$, any local 
jungle equilibrium $\Bz \in \BZ^*(\bthetaJ,h)$ is Pareto 
efficient at $\bthetaJ$.
\end{proposition}

\begin{proof}
Suppose $\Bz \in \BZ^*(\bthetaJ,h)$ is not Pareto efficient
at $\bthetaJ$.
Since this trade profile is feasible for $\bthetaJ$, there must
be another trade profile $\hat \Bz \in \BZ (\bthetaJ)$ such
that $v_i (\hat z_i,\thetaJ_i) \, \geq \, v_i (z_i,\thetaJ_i)$ 
for all $i \in \cali J$, with strict inequality for at 
least some individual in the jungle economy. 

Let $h(n)$ be the most powerful individual in the hierarchy 
$h$ with 
	\begin{equation*}
		v_{h(n)} \big(\hat z_{h(n)}, \thetaJ_{h(n)}\big) 
		\ > \ v_{h(n)} \big(z_{h(n)}, \thetaJ_{h(n)} \big) .
	\end{equation*} 
Hence, $v_{h(m)} \big(\hat z_{h(m)},\thetaJ_{h(m)} \big) 
\, = \, v_{h(m)} \big(z_{h(m)},\thetaJ_{h(m)}\big)
\, = \, \pi_{h(m)}(\bthetaJ)$, for all $1 \leq m < n$.
It follows that $\hat\Bz$ belongs to $\BZ_{h(n-1)} (\bthetaJ)$, 
which is the constraint set of $h(n)$'s optimization problem.  
Immediately, $v_{h(n)} \big(\hat z_{h(n)}, \thetaJ_{h(n)}\big) 
\, \leq \, \pi_{h(n)} (\bthetaJ) \, = \, v_{h(n)} 
\big( z_{h(n)},\thetaJ_{h(n)})$, a contradiction.
\end{proof}

\smallskip

In obtaining the efficiency of the jungle equilibria in
$\BZ^*(\bthetaJ,h)$, we do not require any monotonicity 
assumption on preferences for commodities.  
Without monotonicity, an individual in the jungle may be 
better off disposing of one (or more) goods.  
The exact clearing condition in \autoref{D:feasible-trades}(b) 
avoids this.  
Feasibility could be weakened to the following: 
a trade profile $\Bz$ is ``feasible'' at $\bthetaJ$
if $z_j \in Z_j(\thetaJ_j)$ for all $j \in \cali J$ and
$\sum_{j \in \cali J} z_j \leq \zero$.
This weakened feasible set is still nonempty and compact, so
\autoref{P:local-j-existence} and
\autoref{P:local-j-efficiency} follow verbatim, without any
monotonicity assumption.
Monotonicity is relevant only if one wants the weak inequality
to bind, a \emph{no-waste} reading in which disposal is never
desirable or imposes additional costs on individuals or on
society.
Monotonicity alone does not deliver this: the set of feasible
trades $Z_j(\thetaJ_j)$ is bounded, so an individual who
strictly prefers more of a disposed good $\ell$ may have no room
to absorb this.

The local jungle equilibria in $\BZ^*(\bthetaJ,h)$ are 
efficient, and thus undominated; we focus on them for the 
remainder of the paper. 
In \autoref{S:app-payoff-dom} we show that, under an 
additional assumption, which we call hierarchical improvement, 
any equilibrium in $\BZ^*(\bthetaJ,h)$ weakly payoff 
dominates every other local equilibrium in the jungle.
In their setting without production, \citet{Piccione:2007nh} 
obtain a condition of this kind from primitives. 
They call a jungle \emph{smooth} when each utility
function is strictly quasi-concave, continuously differentiable, 
with strictly positive gradient vector, and each consumption set
has the form $\left\{x \sth g^i(x) \le 0\right\}$ with $g^i$
quasi-convex and differentiable, with $\nabla g^i \gg \zero$. 
In a smooth jungle, whenever an individual can improve upon 
a feasible allocation while keeping stronger
individuals' bundles fixed, she can do so by taking 
unallocated goods or goods from a single 
weaker individual, and the jungle equilibrium is unique.  
Their argument does not carry over to our setting, for two
reasons. 
First, it relies on strict monotonicity of preferences, which 
we do not impose. 
Second, their notion of feasibility allows free disposal: 
the victim remains feasible after surrendering what the deviant 
takes, and whatever the deviant releases can simply be 
discarded. 
Under the exact clearing of \autoref{D:feasible-trades}(b), 
by contrast, a bilateral deviation requires the victim 
to absorb what the deviant releases.  
More generally, any released commodities must be 
accommodated through a feasible reshuffle that preserves 
the payoffs of third parties.
\autoref{S:app-payoff-dom} also shows that, under 
hierarchical improvement, strengthening the concavity of 
$v_i(\cdot,\thetaJ_i)$ to strict concavity delivers a 
unique local jungle equilibrium.

\subsection{Random Hierarchies and Occupational Equilibria}
		\label{SS:random-power}

Our concept of local jungle equilibrium has been stated
for a \emph{fixed} occupational profile $\bthetaJ \in \BThetaJ$ 
and a power relation $h \in \scr H(\cali J)$.
Any equilibrium trade profile $\Bz^* \in \BZ^*(\bthetaJ,h)$
generates the same payoff to individual $h(n) \in \cali J$, 
namely  $\pi_{h(n)} (\bthetaJ)$. 
A main goal in our work is to consider the coexistence of the 
market and the jungle.  
To investigate this, we study occupational choice
and outline a further extension to participation decisions.
First, we let individuals in the jungle decide which 
occupations to choose; i.e., each individual $i \in \cali J$ 
will optimally choose an occupation in $\ThetaJ_i$, 
the set of occupations available for her in the jungle. 
(Later, we do the same for the market economy.) 
A further extension would let individuals 
decide whether to join the jungle or the market 
in anticipation of the resulting occupational choices.
A stable situation for our society, i.e., 
a \emph{social equilibrium}, would endogenize the size 
and composition of these two economies.

Here we focus on the first extension and consider how 
those individuals who are already in the jungle choose 
occupations optimally.  
Since payoffs in a local jungle equilibrium depend on 
the hierarchy, it is entirely plausible that individuals 
would invest in achieving higher rankings in the jungle 
hierarchy.  
\citet{Skaperdas:1992gd} studies a static model in which
individuals allocate limited resources between the cooperative
production of a consumption good and ``arms''.
We do not consider individuals fighting their way to the top of
the hierarchy in the jungle, taking instead the power relation 
as random.%
		\footnote{\citet{Cortes-McPherson:2019pv} recounts 
		a case from the small-scale gold mining area of 
		Madre de Dios, Peru: Gregoria Casas Huamanhuillca, 
		``la Reina del Oro,'' a Quechua-speaking woman 
		from the Peruvian highlands who became a key 
		player in the region.}
Random hierarchies capture uncertainty about 
the exercise of coercive power.  
This interpretation is consistent with research on 
conflicts within authoritarian ruling coalitions 
\citep{Svolik:2009in, Acemoglu:2008lo} and on stochastic 
changes in the distribution of political power 
\citep{Acemoglu:2015po}.  
We represent this uncertainty by assuming that occupational 
choices precede the realization of the hierarchy 
governing allocation.

Still working with a fixed jungle $\cali J \ne \emptyset$,
recall the set of hierarchies $\scr H(\cali J)$
contains all bijections $h \colon \left\{1,2,\ldots, 
|\cali J|\right\} \to \cali J$.  
Since the jungle $\cali J$ is finite, there are finitely
many bijections.  
Impose a uniform distribution on $\scr H(\cali J)$, so that
each power relation is chosen with probability
$1/|\cali J|!$.
In our framework, individuals anticipate that joining the
jungle will give them an equal chance to occupy a certain
position in the jungle hierarchy. 
For most of our results, this is a normalization;  
we only require that individuals not choose 
their positions in the hierarchy themselves: any fixed 
distribution on $\scr H(\cali J)$ will do. 
What is important is that the ranking in the jungle hierarchy is 
not known when occupations are chosen, and does not depend on 
the resources an individual brings to the jungle. 
Uniformity of chances just adds symmetry.

Once individuals in the jungle choose occupations and 
$\bthetaJ \in \BThetaJ$ is fixed, each $i \in \cali J$ obtains 
in expectation a \emph{jungle payoff} equal to 
	\begin{equation}  \label{Eq:expec-payoff-jungle}
		\bar\pi_{i} (\bthetaJ)
		\ = \ \E_{\scr H(\cali J)} \left[ \pi_{h^{-1}(i)} 
			(\bthetaJ) \right]
		\ = \ \frac{1}{|\cali J|!} \sum_{h \in \scr H(\cali J)}
			\pi_{h^{-1}(i)}(\bthetaJ) .
	\end{equation} 
All individuals in the jungle economy anticipate 
that, given an occupational profile $\bthetaJ$ and hierarchy 
a $h$, a local jungle equilibrium in $\BZ^*(\bthetaJ,h)$ 
is selected. 
Every equilibrium in this family provides individual $i$ 
with the payoff $\pi_{h^{-1}(i)}(\bthetaJ)$ when the 
prevailing hierarchy is $h \in \scr H (\cali J)$.%
	\footnote{We slightly abuse notation here and write 
		$\pi_{h^{-1}(i)}(\bthetaJ)$ for the equilibrium 
		payoff of individual $i$ at the given occupational 
		profile, when her ranking under the hierarchy $h$ 
		is $h^{-1}(i)$.}  
This is the expression inside the expectations
operator in \autoref{Eq:expec-payoff-jungle}.  
Averaging over all possible hierarchies (under the 
uniform distribution) obtains $i$'s jungle 
payoff $\bar\pi_{i} (\bthetaJ)$.

An occupational equilibrium in the jungle economy is a 
profile of occupational choices that gives no individual 
incentives to deviate unilaterally.

\begin{definition}  \label{D:jungle-occup-equi}
An \emph{occupational equilibrium in the jungle $\cali J$}
is a profile $\bsthetaJ = (\sthetaJ_i) \in \BThetaJ$ 
such that, for each individual $i \in \cali J$ and each 
alternative occupation $\thetaJ_i \in \ThetaJ_i$,
	\begin{equation}  \label{Eq:jungle-OE}
		\bar\pi_i(\sthetaJ_i,\bsthetaJ_{-i})
		\ \ge \ \bar\pi_i(\thetaJ_i,\bsthetaJ_{-i}) .
	\end{equation}
\end{definition}

\subsection{Existence of Occupational Equilibria in the 
Jungle}  \label{S:occup-existence} 

An occupational equilibrium in the jungle is a (pure) Nash 
equilibrium of the game in which individuals simultaneously 
choose occupations and anticipate the expected payoffs 
from the jungle allocation. 
Even when the occupation sets are compact subsets of 
Euclidean spaces, the expected payoff function in 
\autoref{Eq:expec-payoff-jungle} need not be continuous, 
so neither Berge's Maximum Theorem nor standard fixed point 
arguments apply to the occupational game in the jungle, 
as it stands.
Hence, we cannot take existence of an occupational equilibrium 
for granted.

There are two issues in proving existence of occupational 
equilibria in the jungle.  
The first, common to parametric optimization problems, is 
that individual $i$'s occupational choice affects not only 
her utility function but also the set of individually 
feasible trade positions she has at her disposal in the 
jungle economy.  
To deal with this, we exploit the structure given by 
\autoref{A:structure} to the set of occupational choices 
$\ThetaJ_i$, the correspondence $Z_i \colon \ThetaJ_i 
\rightrightarrows \R^L$, and the valuation function 
$v_i$.

The second issue is novel and comes from the serial recursion 
used to obtain existence of local jungle equilibria in 
\autoref{P:local-j-existence}. 
In this recursion, the constraint set at each stage, except for  
the initial one, is the solution set of the preceding stage. 
Thus, when treated as a correspondence, the constraint 
correspondence of all but the initial step in our nested 
optimization problem is endogenous, and cannot simply be 
assumed to be a continuous correspondence. 
As a result, this nesting need not deliver continuity of 
the individual value functions. 
The next examples illustrate this issue and point 
towards ways to deal with it.

\begin{example}  \label{Ex:serial-jump}
There is only one commodity in this economy, so $L=1$. 
The jungle is populated by two individuals, $\cali J = \{1,2\}$, 
with $\ThetaJ_1 = [0,1]$ and $\ThetaJ_2 = \{1\}$, so only the 
first individual has occupational choices. 
For both $i = 1,2$, let $Z_i(\thetaJ_i) = [-1,1]$ and 
$v_i(z,\thetaJ_i) = \thetaJ_i \, z$ on this interval, with value 
$-\infty$ outside it. 
Clearly, all of the conditions in \autoref{A:structure} hold 
in this setting. 

For simplicity, express the set of feasible trade profiles 
in this jungle economy as the line segment $\BZ = \left\{ t 
\sth t \in [-1,1]\right\}$, 
with the understanding that $t \in \BZ$ allocates quantity $t$ 
to individual $1$ and quantity $-t$ to individual $2$; 
we omit dependence of $\BZ$ on the occupational profile 
since each $Z_i(\thetaJ_i)$ does not depend on $\thetaJ_i$.   

Consider the hierarchy $h \in \scr H(\cali J)$ that 
sets $h(1) = 1$. 
Clearly, the first dictator chooses $t = 1$ if 
$\thetaJ_{h(1)} > 0$, but at the occupational choice 
$\thetaJ_{h(1)} = 0$, she is indifferent over the 
entire $\BZ$.  
Thus, $h(1)$'s solution correspondence in her optimization 
problem, which is the feasible correspondence in $h(2)$'s 
optimization problem, is 
	\begin{equation*}
		\BZ_{h(1)}(\bthetaJ) 
		\ = \ 
		\begin{cases}
			[-1,1] 	\ 
			  & \colon \, \thetaJ_{h(1)} \, = \, 0 , \\[.75ex]
			\{1\} \ 
			  & \colon 0 \, < \, \thetaJ_{h(1)} \, \leq \, 1 .
		\end{cases}
	\end{equation*}
This correspondence is upper hemicontinuous but not lower 
hemicontinuous. 
Readily, 
	\begin{equation*}
		\pi_{h(1)}(\bthetaJ) \ = \ \thetaJ_{h(1)} 
			\andeq 
		\pi_{h(2)} (\bthetaJ) \ = \
		\begin{cases}
			1 \  
			  & \colon \, \thetaJ_{h(1)} \, = \, 0 , \\[.75ex] 
			- \, 1 \ 
			  & \colon \, 0 \, < \, \thetaJ_{h(1)} \, \leq \, 1 .
		\end{cases}
	\end{equation*}

Thus, with $h(1) = 1$ at the top of the hierarchy, individual
$1$'s payoff function coming from local jungle equilibria 
is continuous, 
but individual 2's payoff function exhibits a jump at 
$\thetaJ_1 = 0$.  
Under the reverse order, individual 2 chooses $t = -1$ and 
obtains 1, and individual 1 obtains $- \thetaJ_1$.
But the averaging over the two power relations does not repair
the discontinuity in the expected payoff function of the second 
individual: $\bar\pi_2 (\bthetaJ) = 1$ for $\thetaJ_1 = 0$ and 
$\bar\pi_2 (\bthetaJ)=0$ for all $\thetaJ_1 > 0$.  
\hfill $\diamond$ 
\end{example}

The discontinuity in the jungle payoff function of individual 
$2$ in \autoref{Ex:serial-jump} is generated by the 
multiplicity of solutions at one of individual $1$'s 
occupational choices.  
Replacing concavity of the valuation function with strict 
concavity makes the optimizing individual's trade unique 
at each stage, but need not determine the other individuals' 
trades.  
Because of this latter point, it is not sufficient to 
guarantee the continuity of payoffs. 
The next example illustrates this new difficulty. 
All individual valuations are strictly concave, and the 
most powerful individual's optimal trade is unique and 
does not change with occupational choices. 
Nevertheless, a weaker individual's jungle payoff 
turns out to be discontinuous.

\begin{example}  \label{Ex:strict-convexity}
There are two goods in this economy, so $L = 2$.  
The jungle is now populated by three individuals, 
$\cali J = \{1,2,3\}$, with $\ThetaJ_1 = \ThetaJ_2 = \{1\}$ 
and $\ThetaJ_3 = [0,1]$.  
Thus, only the third individual has occupational choices. 
Let
	\begin{align*}
		Z_1(1) \ = \ [-1,1]^2,  \qquad
		Z_2(1) \ = \ [-1,0] \times [0,1],  
			\qquad \text{and}	\\[.75ex]
		Z_3(\thetaJ_3)
		\ = \ \big\{ (z_{31},z_{32}) \,\in\, [-1,0] \times [0,1]  
		\sth \thetaJ_3 z_{31}  \, + \, z_{32} \ \geq \ 0 \big\} .
	\end{align*}
Individual $2$ can supply up to one unit of the first good 
without using the second good as an input.  
Individual $3$'s occupation determines her input requirement: 
selling one unit of the first good requires at least 
$\thetaJ_3$ units of the second.
On the respective individual feasible sets, let valuations 
be defined by 
	\begin{align*}
		v_1(z_1,1)
		& \ = \ 2z_{11} \, - \, z_{11}^2 \, - \, z_{12}^2,  \\
		v_i(z_i,\thetaJ_i)
		& \ = \ - \, z_{i1}^2 \, - \, z_{i2}^2,
			\qquad \text{for $i = 2,3$,}
	\end{align*}
with value $-\infty$ outside those sets. 
Notice that individual $1$ prefers to receive one unit of 
the first good and not to trade the second. 
Individuals $2$ and $3$ bear the cost when required to produce.

All conditions in \autoref{A:structure} are present. 
In particular, the correspondence $Z_3$ is continuous.  
Moreover, since zero belongs to both $Z_2(1)$ and 
$Z_3(\thetaJ_3)$, their aggregate with $Z_1(1)$ contains 
$[-1,1]^2$, and thus $\zero$ is in its interior. 
Notice the valuation functions are not only continuous on 
their feasible graphs but also strictly concave in 
trade.  
The individual feasible sets, however, are not strictly 
convex.

Consider the hierarchy with $h(1) = 1$, $h(2) = 2$ and 
$h(3) = 3$.  
The strongest individual's  unique optimal trade is $(1,0)$.  
This trade is feasible at every occupational profile. 
Thus individual $h(1) = 1$ always chooses $(1,0)$ and 
obtains payoff $1$.  
Because the second coordinates of the remaining individuals' 
feasible trades must be nonnegative, the clearing condition 
now requires $z_{22} = z_{32} = 0$.
If $\thetaJ_3 > 0$, individual $3$'s input requirement forces 
$z_{31} = 0$. 
In this case, individual $2$ must supply the entire unit of 
the first commodity consumed by the first dictator.  
If instead $\thetaJ_3 = 0$, either remaining individual can 
supply it.  
The first dictator's solution correspondence is therefore 
	\begin{equation*}
		\BZ_{h(1)}(\bthetaJ)
		\ = \ 
		\begin{cases}
		\left\{ \bigl( (1,0),(-t,0),(t-1,0) \bigr) 
			\sth t \,\in\, [0,1] \right\} ,
			& \thetaJ_3\ = \ 0,  \\[1ex]
		\left\{ \bigl( (1,0),(-1,0),(0,0) \bigr) \right\}, 
			& 0 \ < \ \thetaJ_3 \ \leq \ 1.
		\end{cases}
	\end{equation*}
This correspondence is upper hemicontinuous but not 
lower hemicontinuous at $\thetaJ_3 = 0$.  
At this occupation, individual $h(2) = 2$ chooses $t = 0$ 
and leaves all production to individual $h(3) = 3$.  
Consequently,
	\begin{equation*}
		\pi_{h(2)}(\bthetaJ)
		\ = \ 
		\begin{cases}
			0,  & \thetaJ_3 \ = \ 0,  \\[.75ex]
			- \, 1,  & 0 \ < \ \thetaJ_3 \ \leq \ 1.
		\end{cases}
	\end{equation*}

Averaging over hierarchies does not repair the 
discontinuity in individual $2$'s payoff.  
In the power orders $(1,3,2)$ and $(3,1,2)$, listed from 
strongest to weakest, individual $3$ chooses $\zero$ and 
individual $1$ takes $(1,0)$, leaving individual $2$ 
with $(-1,0)$ and a payoff of $-1$.  
In the orders $(2,1,3)$, $(2,3,1)$ and $(3,2,1)$, individual 
$2$ chooses $\zero$ and obtains payoff $0$. 
These five payoffs do not depend on $\thetaJ_3$, and hence 
$\bar\pi_2(\bthetaJ) = - 1/3$ at $\thetaJ_3 = 0$ and 
$\bar\pi_2(\bthetaJ) = - 1/2$ for all $0 < \thetaJ_3 \leq 1$. 
\hfill $\diamond$
\end{example}

To spell out the issue illustrated by these examples, 
fix a hierarchy $h$ in the jungle $\cali J$.  
For each position $n = 1,2,\ldots,|\cali J|$, the correspondence 
$\BZ_{h(n)} \colon \BThetaJ \rightrightarrows \R^{|\cali J| 
\times L}$ employed in the proof of 
\autoref{P:local-j-existence} can be expressed as 
	\begin{equation}  \label{Eq:level-set-form}
		\BZ_{h(n)}(\bthetaJ)
		\ = \ \left\{ \Bz\in\BZ_{h(n-1)}(\bthetaJ) \sth 
			v_{h(n)} \big(z_{h(n)},\thetaJ_{h(n)}\big)
			\, = \, \pi_{h(n)}(\bthetaJ) \right\}.
	\end{equation} 
Continuity of $\pi_{h(n)}(\bthetaJ)$ on $\BThetaJ$ for every 
hierarchy $h$ would imply continuity of the expected 
payoff function in \autoref{Eq:expec-payoff-jungle}, 
since it is an averaging over finitely many power orders.  
But $\pi_{h(n)}$ is the value function coming from individual 
$h(n)$'s optimization problem, so one is required to show, not 
assume, its continuity.  
Moreover, $\BZ_{h(n-1)}$ is the constraint correspondence 
at stage $n$ in the recursion but the solution correspondence 
of the previous stage, and thus it need not be lower 
hemicontinuous even when it is upper hemicontinuous.

\autoref{Ex:serial-jump} illustrates the problems created 
when a stage in the nested optimization problem of 
our serial dictatorship procedure has multiple solutions.  
The difficulty of \autoref{Ex:strict-convexity} has 
a different source: it relates to how weaker individuals 
can accommodate a stronger individual's chosen trade, even
when that optimal trade is unique.  
To deal with these difficulties, we strengthen the 
structure of the jungle economy by imposing strict 
concavity of valuations on trade and strict 
convexity of individual feasible sets.   
We next show that these stronger conditions 
are sufficient to guarantee that the nested optimization 
problem in \autoref{P:local-j-existence} has, at each stage, 
a constraint correspondence that is not only upper 
hemicontinuous but also lower hemicontinuous.

\begin{proposition}   \label{P:correspondence-j-continuity}
Suppose that, complementing the conditions provided in 
\autoref{A:structure} as applied to the jungle, for all 
$i \in \cali J$ the correspondence $Z_i \colon \ThetaJ_i 
\rightrightarrows \R^L$ has strictly convex values%
	\footnote{That is, for any two distinct $z_i,z'_i \in 
		Z_i(\thetaJ_i)$, for any $\lambda\in(0,1)$, 
		the trade position $\lambda z_i + (1-\lambda) z'_i
		\in \interior_{\R^L} Z_i(\thetaJ_i)$. 
		This allows $Z_i(\thetaJ_i)$ to be a singleton, 
		e.g., $Z_i(\thetaJ_i) = \{\zero\}$, 	but 
		excludes non-degenerate, lower dimensional 
		individual trade sets. }
in $\R^L$ and the valuation function
$v_i(\cdot,\thetaJ_i)$ is strictly concave on 
$Z_i(\thetaJ_i)$, for all $\thetaJ_i \in \ThetaJ_i$. 
Given any hierarchy $h \in \scr H(\cali J)$, the following 
hold: 
\begin{enumerate}[leftmargin=1.5em,label=(\alph*)] 
		\setcounter{enumi}{0}  \itemsep.25ex
	\item For every $n = 0,1,\ldots,|\cali J|$, the 
		correspondence $\bthetaJ \rightrightarrows 
		\BZ_{h(n)}(\bthetaJ)$ defined in 
		\autoref{Eq:level-set-form} is continuous and has 
		nonempty, compact and convex values.  
	\item For every $n = 1,\ldots,|\cali J|$, the 
		payoff function $\bthetaJ \mapsto \pi_{h(n)} 
		(\bthetaJ)$ is continuous on $\BThetaJ$. 
\end{enumerate}

\end{proposition}  

\begin{proof}
Fix a hierarchy $h \in \scr H(\cali J)$ for the rest 
of the proof.
By \autoref{L:feasible-trades} and 
\autoref{P:local-j-existence}, for any occupational profile 
$\bthetaJ$, the sets $\BZ_{h(n)}(\bthetaJ)$ are nonempty 
and compact at every stage $n = 0,1,\ldots, |\cali J|$. 
Their convexity follows from the argument in 
\autoref{C:local-j-properties}(a). 

For $n \geq 1$, strict concavity of the valuation function 
implies that any two trade profiles in $\BZ_{h(n)}(\bthetaJ)$ 
agree in the individual trade positions of 
$h(1),h(2),\ldots,h(n)$. 
Indeed, if two maximizers at stage $n$ in the recursion 
differed in individual $h(n)$'s position, their midpoint 
would belong to the convex set $\BZ_{h(n-1)}(\bthetaJ)$ 
and give this individual strictly more than $\pi_{h(n)} 
(\bthetaJ)$, which is impossible.  
The same argument at the preceding stages, together with 
the nesting of the constraint sets in the proof of 
\autoref{P:local-j-existence}, gives agreement in all 
earlier positions.

We claim that, at each stage, the upper hemicontinuity of 
the correspondence $\BZ_{h(n)}$ implies its lower 
hemicontinuity.  
To see this, fix some $n \in \{0,1,\ldots,|\cali J|\}$ and 
suppose that $\BZ_{h(n)}$ is upper hemicontinuous 
at occupation profile $\bthetaJ$.  
Consider any sequence $\btheta^r \to \bthetaJ$ in $\BThetaJ$
and any target $\Bz \in \BZ_{h(n)}(\bthetaJ)$.  
For each $r = 1,2,\ldots$, choose $\Bz^r \in \BZ_{h(n)} 
(\btheta^r)$ with minimum distance to $\Bz$. 
Such trade profiles exist by the nonemptiness and compactness 
of the sets $\BZ_{h(n)} (\btheta^r)$.  

From \autoref{A:structure}(b), the individual feasible 
correspondences are locally bounded.  
Thus, if the sequence $\{\Bz^r\}$ does not converge to $\Bz$, 
the upper hemicontinuity of $\BZ_{h(n)}$ at $\bthetaJ$ 
gives a subsequence, which we still index by $r$, such that 
$\Bz^r \to \hat{\Bz} \in \BZ_{h(n)}(\bthetaJ)$ and 
$\hat{\Bz} \neq \Bz$.
Along this subsequence, define
	\begin{equation*}
		\tilde{\Bz}^{r} \ = \ 
		\Bz^r \ + \ \tfrac{1}{2}(\Bz \, - \, \hat{\Bz}) .
	\end{equation*}
If $z_i = \hat z_i$ for $i \in \cali J$, then $\tilde z_i^r 
= z_i^r \in Z_i(\theta_i^r)$.   
If instead $z_i \neq \hat z_i$, strict convexity of the 
individual trading sets implies 
	\begin{equation*}
		\tilde z_i^r \ \to \ \frac{z_i \ + \ \hat z_i}{2}
		\ \in \ \interior \, Z_i(\thetaJ_i).
	\end{equation*}
The continuity and convexity of the correspondence 
$Z_i$ on $\ThetaJ_i$ imply that a fixed neighborhood of 
this midpoint is contained in $Z_i(\theta_i^r)$, for all 
sufficiently large $r$.  
To see this, choose a simplex contained in $Z_i(\thetaJ_i)$ 
with the midpoint in its interior.  
Lower hemicontinuity allows the vertices of this simplex 
to be approximated by points in $Z_i(\theta_i^r)$. 
By convexity, their convex hulls are contained in 
$Z_i(\theta_i^r)$ and still contain a fixed neighborhood of
the midpoint. 
Consequently, $\tilde z_i^r \in Z_i(\theta_i^r)$ for all 
sufficiently large $r$.

Since there are finitely many individuals, all 
trading positions in $\tilde{\Bz}^r$ are individually 
feasible, again for sufficiently large $r$. 
Moreover,
	\begin{equation*}
		\sum_{i \in\cali J} \tilde z_i^r
		\ = \ \sum_{i\in\cali J} z_i^r \ + \ \tfrac{1}{2} 
			\Bigl( \sum_{i \in \cali J} z_i \, - \, 
			\sum_{i \in \cali J} \hat z_i \Bigr)
		\ = \ \zero.
\end{equation*}
Hence $\tilde{\Bz}^r \in \BZ(\btheta^r)$.  
For $n \geq 1$, the agreement of $\Bz$ and $\hat{\Bz}$ in the
first $n$ positions in the hierarchy also shows that 
	\begin{equation*}
		\tilde z_{h(m)}^r \ = \ z_{h(m)}^r,
			\qquad \text{for all $m = 1,\ldots,n$.}  
	\end{equation*}
These equalities imply that $\tilde{\Bz}^r$  
attains the same payoff as $\Bz^r$ at each of the first 
$n$ nested optimization problems.  
Thus $\tilde{\Bz}^{r} \in \BZ_{h(n)} (\btheta^r)$ for all
sufficiently large $r$.  
Note this also holds when $n = 0$.
Now, the choice of $\Bz^r$ as a point of minimum 
distance to $\Bz$ gives  
	\begin{equation*}
		\|\Bz^r \, - \, \Bz\|
		\ \leq \ \|\tilde{\Bz}^r \, - \, \Bz\|.
	\end{equation*}
Taking limits along the subsequence obtains 
	\begin{equation*}
		\|\hat{\Bz} \, - \, \Bz\|
		\ \leq \ \tfrac{1}{2} \, \|\hat{\Bz} \, - \, \Bz\|,
	\end{equation*}
which contradicts $\hat{\Bz}\neq\Bz$. 
Therefore $\Bz^r \to \Bz$, which shows that the 
lower hemicontinuity of $\BZ_{h(n)}$ at $\bthetaJ$ 
follows from its upper hemicontinuity at $\bthetaJ$. 

We are ready to establish the continuity properties of 
(a) and (b) in the proposition.

\smallskip
\noindent \emph{Initial Step.} 
The individual feasible correspondences are locally bounded
and have closed graphs. 
The feasibility constraint set $\BF = \{\Bz \sth  
\sum_{i \in \cali J} z_i = \zero\}$ is also closed. 
Therefore, the correspondence $\BZ_{h(0)} = \BZ$ defined 
on $\BThetaJ$ is upper hemicontinuous. 
The preceding argument gives lower hemicontinuity, so 
$\BZ_{h(0)}$ is continuous.

\smallskip
\noindent \emph{Induction Step.} 
For $1 \leq n \leq |\cali J|$, suppose that $\BZ_{h(n-1)}$
is a continuous correspondence. 
Individual $h(n)$'s optimization problem can be expressed as 
	\begin{equation*}
		\pi_{h(n)}(\bthetaJ)
		\ = \ \max_{\Bz} \left\{ v_{h(n)} \big(z_{h(n)}, 
			\thetaJ_{h(n)} \big) \sth \Bz \, \in \, 
			\BZ_{h(n-1)}(\bthetaJ) \right\} .
	\end{equation*}
From \autoref{A:structure}(c), the objective function 
is finite and continuous on the graph of the constraint 
correspondence. 
Berge's Theorem of the Maximum implies that $\pi_{h(n)}$ is
continuous and that $\BZ_{h(n)}$ is upper hemicontinuous.
The preceding argument gives lower hemicontinuity, hence
continuity of $\BZ_{h(n)}$.

This establishes part~(a) for $n = 0,1,\ldots, 
|\cali J|$ and part (b) for $n = 1,\ldots,|\cali J|$.
\end{proof}

We can now state our existence result for the jungle economy. 
In the main body of the paper we focus on pure 
occupational equilibria.  
This will allow us to compare in a relatively simple way 
properties of occupational equilibria in the jungle and  
in the market.  
For that purpose, we assume that any convex combination 
of best replies is also a best reply.    
In \autoref{S:app-existence}, we consider existence of 
mixed occupational equilibria, which do not rely on 
any additional assumption.

\begin{theorem}  \label{T:jungle-occup-pure} 
Let the conditions of \autoref{P:correspondence-j-continuity} 
be in place. 
Suppose in addition that, for every individual $i \in\cali J$ 
and every $\bthetaJ_{-i} \in \times_{j \neq i} \ThetaJ_j$, 
the set of maximizers of $\thetaJ_i \mapsto \bar\pi_i 
(\thetaJ_i,\bthetaJ_{-i})$ over $\ThetaJ_i$ is convex. 
Then an occupational equilibrium in the jungle, in the 
sense of \autoref{D:jungle-occup-equi}, exists.
\end{theorem}

\begin{proof}
We know, from \autoref{P:correspondence-j-continuity}(b), 
that each individual's payoff function is continuous 
in occupational profiles for every fixed hierarchy $h$. 
Since $\scr H (\cali J)$ is finite, the expected payoff 
$\bar\pi_i$ is continuous on $\BThetaJ$, for every 
$i\in\cali J$.

From \autoref{A:structure}(a), as imposed on the jungle, 
$\BThetaJ = \times_{i\in\cali J} \ThetaJ_i$ is a nonempty, 
compact and convex subset of a finite-dimensional normed 
vector space. 
For each individual $i$, define the best reply correspondence 
	\begin{equation*}
		\BR_i(\bthetaJ_{-i})
		\ \equiv \ \argmax \, \big\{ \bar\pi_i(\thetaJ_i,
			\bthetaJ_{-i}) \sth \thetaJ_i \in \ThetaJ_i \big\} .
		\end{equation*}
Berge's Theorem of the Maximum implies that $\BR_i$ is 
upper hemicontinuous and has nonempty, compact values.  
The additional assumption in the theorem makes these values 
convex.  
Hence the product correspondence
	\begin{equation*}
		\BR(\bthetaJ)
		\ \equiv \ \times_{i \in \cali J} 
			\, \BR_i(\bthetaJ_{-i})
	\end{equation*}
maps $\BThetaJ$ into itself, is upper hemicontinuous, and
has nonempty, compact and convex values.  
Applying Kakutani's Fixed Point Theorem obtains an 
occupational profile $\bsthetaJ \in \BThetaJ$ such that 
$\bsthetaJ \in \BR(\bsthetaJ)$.  
Thus, for every $i\in\cali J$ and every alternative occupation 
$\thetaJ_i\in\ThetaJ_i$, we have 
	\begin{equation*}
		\bar\pi_i(\sthetaJ_i,\bsthetaJ_{-i})
		\ \geq \ \bar\pi_i (\thetaJ_i,\bsthetaJ_{-i}).  
		\qedhere 
	\end{equation*}
\end{proof}

The existence argument in \autoref{T:jungle-occup-pure} 
follows the standard best reply fixed-point approach of 
\cite{Debreu:1952ki}, \cite{Fan:1952wa} and 
\cite{Glicksberg:1952vp}.  
Our \autoref{P:correspondence-j-continuity} 
establishes continuity of the expected occupational payoffs 
generated by the serial dictatorship in the jungle.  
Given this continuity, compact and convex occupational menus 
and convex sets of occupational best replies permit an 
application of Kakutani's Fixed Point Theorem. 
The extra assumption for our existence of a pure occupational 
equilibrium in the jungle in \autoref{T:jungle-occup-pure} 
restricts an endogenous object. 
The following special case provides a sufficient condition 
on primitives.

\begin{corollary}   \label{C:jungle-pure-primitive} 
Let the conditions of \autoref{P:correspondence-j-continuity} 
be in place. 
Assume that, for every individual $i \in\cali J$ in the 
jungle, the feasible trade set $Z_i(\thetaJ_i)$ is independent 
of $\thetaJ_i$, and denote it simply by $Z_i$.  
If the function $(z_i,\thetaJ_i) \mapsto v_i 
(z_i,\thetaJ_i)$ is jointly concave on $Z_i \times \ThetaJ_i$, 
for $i \in \cali J$, then a pure occupational equilibrium in 
the jungle exists.
\end{corollary}

\begin{proof}
Fix an individual $i \in \cali J$ and a hierarchy $h$, and  
let $n$ be the position for which $h(n) = i$.  
Fix also an occupational profile $\bthetaJ_{-i}$ for the 
other individuals.
Because the individual feasible trade sets are independent
of occupations, changing $\thetaJ_i$ does not change the
initial feasible set $\BZ = \BZ_{h(0)}$.   
It also does not change the objective functions of the 
individuals preceding $i$ in the hierarchy.  
Applying the recursion in \autoref{P:local-j-existence} 
successively, the constraint set $\BZ_{h(n-1)} 
(\thetaJ_i,\bthetaJ_{-i})$ is therefore independent 
of $\thetaJ_i$. 
Moreover, this  set is nonempty, compact and convex.

Consider two occupations $\thetaJ_i, \hthetaJ_i \in 
\ThetaJ_i$, and choose optimal trade profiles 
	\begin{equation*}
		\Bz \, \in \, \BZ_{h(n)}(\thetaJ_i,\bthetaJ_{-i}) 
			\andeq
		\hat{\Bz} \, \in \, \BZ_{h(n)}(\hthetaJ_i, 
			\bthetaJ_{-i}).
	\end{equation*}  
By our previous observation, both $\Bz$ and $\hat{\Bz}$  
belong to the same convex constraint set.  
For every $\lambda \in [0,1]$, their convex combination 
$\lambda\Bz \, + \, (1-\lambda) \hat{\Bz}$ is consequently 
feasible in individual $h(n) = i$'s optimization problem at
occupation $\lambda \thetaJ_i + (1-\lambda)\hthetaJ_i$.  
Joint concavity of the valuation function now implies  
	\begin{align*}
		\pi_{h(n)} \bigl( \lambda\thetaJ_i \, + \, 
			(1-\lambda)\hthetaJ_i,\bthetaJ_{-i} \bigr) 
		& \ \geq \ v_i \bigl(\lambda z_i \, + \, 
			(1 - \lambda)\hat z_i , \, \lambda\thetaJ_i 
			\, + \, (1-\lambda)\hthetaJ_i\bigr)   \\[.75ex] 
		& \ \geq \ \lambda v_i(z_i,\thetaJ_i) \ + \ 
			(1-\lambda) v_i(\hat z_i,\hthetaJ_i)  \\[.75ex]
		& \ = \ 	\lambda \pi_{h(n)}(\thetaJ_i,\bthetaJ_{-i})
			\ + \ (1-\lambda) \pi_{h(n)}(\hthetaJ_i, 
			\bthetaJ_{-i}).
		\end{align*}
Thus individual $i$'s payoff under the hierarchy $h$ 
is concave in her own occupation.  
This holds for every hierarchy in $\scr H (\cali J)$,  
and thus averaging over power relations preserves the 
inequality. 
We conclude that the expected payoff $\bar\pi_i (\cdot, 
\bthetaJ_{-i})$ is concave, and thus its 
set of maximizers is convex.
\autoref{T:jungle-occup-pure} thus applies. 
\end{proof}


\section{The Market}   \label{S:market}

A major difference between the market and the jungle is in 
the enforcement of contracts and property rights.   
In the market, impersonal transactions are enforced by an 
anonymous, centralized institution that, in addition, 
organizes the supply of the numeraire.  
So, in the market the numeraire is tradable, and individual 
choices are settled by posted prices.


\subsection{Local and Occupational Equilibria in the Market}
	\label{SS:equilibria-market}

Our model of a market economy follows \citet{Makowski:1995cf}; 
see related analysis in \cite{Makowski:1987rf}.  
We continue to impose \autoref{A:structure}, adapting it 
to the market economy, and we work with a nonempty market 
throughout this section.

\begin{definition}   \label{D:market-equilibrium}
Fix an occupational profile $\bthetaM = (\thetaM_i) \in 
\BThetaM$ for the market economy $\cali M$.
A pair $(\Bz,p) \in \BZ(\bthetaM) \times \R^L$ is a 
\emph{market equilibrium at $\bthetaM$} if, for each 
individual $i \in \cali M$ and every $z'_i \in \R^L$,
	\begin{equation*}
		v_i(z_i, \thetaM_i) \, - \, p \cdot z_i
		\ \ge \ v_i(z'_i, \thetaM_i) \, - \, p \cdot z'_i .
	\end{equation*}
\end{definition}

The above inequality is $i$'s optimality condition under 
quasilinear preferences with the price of the numeraire 
normalized to one: she chooses a trade and a
monetary holding to maximize $v_i(z_i',\thetaM_i) + m_i'$ 
subject to the trading budget constraint 
$p \cdot z_i' + m_i' = 0$.  
The market clearing condition for the $L$ commodities is 
embedded in the requirement that $\Bz \in \BZ(\bthetaM)$.  
In equilibrium, the market for the numeraire also clears. 
Notice we do not need to impose the restriction $z'_i \in 
Z_i(\thetaM_i)$, since $v_i(z'_i,\thetaM_i) = - \infty$ 
for any $z'_i$ not in $Z_i(\thetaM_i)$.  
We allow prices of either sign: without monotonicity of preferences
a commodity may be a nuisance, and the restriction $p \ge \zero$
would be an additional hypothesis rather than a conclusion.
But this does not play any role in our analysis.

We sometimes refer to a market equilibrium at $\bthetaM$ as 
a \emph{local market equilibrium}. 
\autoref{A:structure}(d), as imposed on the market economy, 
implies that all commodity markets are open in the 
sense of \citet{Makowski:1995cf}: every commodity $\ell = 
1,\ldots,L$ can be potentially supplied at every 
occupational profile.%
		\footnote{Indeed, if commodity $\ell$ could not be 
		supplied, then $z_{i\ell} \geq 0$ for every 
		$i \in \cali{M}$ and every $z_i\in Z_i(\thetaM_i)$.  
		The aggregate trading set would then lie in the 
		half-space $\{\bar z \in \R^L \sth \bar z_\ell 
		\geq 0\}$, contradicting \autoref{A:structure}(d).} 
Thus, in our market economy occupational choices can 
affect trading possibilities and equilibrium prices, 
and markets are complete.    
Note however that this does not require every commodity 
to be traded in equilibrium.

The next proposition establishes the existence of a local 
market equilibrium.  
Our proof uses a standard supporting-price argument; 
we include it for completeness.%
		\footnote{See for example 
		\citet[Theorem 23.4]{Rockafellar:1970uy}.}
\autoref{A:structure}(d) ensures that, in any local market 
equilibrium, the price of the numeraire is strictly positive, 
so our price normalization implicit in 
\autoref{D:market-equilibrium} is without loss of generality.

\begin{proposition}   \label{P:local-m-existence}
For any occupational profile $\bthetaM\in\BThetaM$, a 
local market equilibrium exists.
\end{proposition}

\begin{proof}
Fix an occupational profile $\bthetaM \in \BThetaM$. 
For each $\bar z \in \sum_{i} Z_i(\thetaM_i)$, define
	\begin{equation*}
		\hat g (\bar z)
		\ \equiv \ \max_{\Bz} \, \Bigl\{ \sum_{i \in \cali{M}} 
			v_i(z_i,\thetaM_i) \sth 	z_i \in Z_i(\thetaM_i) \
			 \text{for all $i \in \cali{M}$}, \ 
			 \sum_{i \in \cali{M}} z_i \, =  \, \bar z
			\Bigr\} .
		\end{equation*} 
It is clear that the maximum is attained for all such 
$\bar z \in \sum_{i} Z_i(\thetaM_i)$, since the constraint 
set is nonempty and compact, and the objective function 
is continuous. 
Convexity of the individual feasible sets and concavity 
of valuations imply that $\hat g$ is concave on 
$\sum_{i} Z_i(\thetaM_i)$. 
Choose $\Bz^* \in \BZ(\bthetaM)$ attaining $\hat g(\zero)$.

Now consider the convex set
	\begin{equation*}
		C \ = \ \Bigl\{ 	(\bar z,\bar m) \, \in \, 
			\sum_{i} Z_i(\thetaM_i) \times \R \sth
			\bar m \ \geq \ \hat g(\zero) \, - \,
			\hat g (\bar z) \Bigr\}.
		\end{equation*}
The point $(\zero,0)$ belongs to the boundary of $C$. 
By the Supporting Hyperplane Theorem, there exists a nonzero 
vector $(\bar p,p_m) \in \R^L \times \R$ such that 
	\begin{equation*}
		\bar p \cdot \bar z \ + \ p_m \bar m
		\ \geq \ 0,
 			\qquad \text{for all $(\bar z,\bar m) \in C$.}
	\end{equation*}
Since $(\zero,1)$ belongs to $C$, we have $p_m \geq 0$. 
If $p_m = 0$, the preceding inequality gives $\bar p \cdot 
\bar z \geq 0$ for all $\bar z \in \sum_{i} Z_i(\thetaM_i)$.  
But from \autoref{A:structure}(d), we have that 
$\zero \in \interior \sum_{i} Z_i(\thetaM_i)$, so this 
requires $\bar p = \zero$, a contradiction. 
Thus $p_m > 0$.  
Normalizing the price of the numeraire to one and writing 
$p^* = \bar p/p_m$, we obtain 
	\begin{equation*}
		\hat g(\bar z)
		\ \leq \ \hat g(\zero) \ + \ p^* \cdot \bar z,
			\qquad \text{for all $\bar z \in \sum_{i} 
			Z_i(\thetaM_i)$.}
	\end{equation*}

Fix any individual $i \in \cali{M}$ and any $z_i' \in 
Z_i(\thetaM_i)$.  
Replacing $z_i^*$ by $z_i'$ in $\Bz^*$ and using the 
preceding inequality gives 
	\begin{equation*}
		v_i(z_i',\thetaM_i) \ + \ \sum_{j \ne i} 
			v_j(z_j^*,\thetaM_j) 
		\ \leq \ \hat g(z_i' - z_i^*) 
		\ \leq \ \hat g(\zero) \ + \ p^* \cdot (z_i'-z_i^*).
		\end{equation*}
Since $\hat g(\zero) = \sum_{j \in \cali{M}} v_j(z_j^*, 
\thetaM_j)$, rearranging the above expression results in 
	\begin{equation*}
		v_i(z_i^*,\thetaM_i) \ - \ p^* \cdot z_i^*
		\ \geq \ v_i(z_i',\thetaM_i) \ - \ p^* \cdot z_i'.
	\end{equation*}
For $z_i' \notin Z_i(\thetaM_i)$, the same inequality 
holds because $v_i(z_i',\thetaM_i) = -\infty$. 
Thus $(\Bz^*,p^*)$ satisfies \autoref{D:market-equilibrium}. 
Finally, the monetary positions $m_i^* = -p^* \cdot z_i^*$  
satisfy $\sum_{i} m_i^* = 0$, so the market for the 
numeraire also clears.
\end{proof}

The following is well-known, e.g., 
\citet[Proposition 1]{Makowski:1995cf}: 
local market equilibria produce efficient allocations.

\begin{proposition}   \label{P:local-m-efficiency} 
If $(\Bz^*,p^*)$ is a market equilibrium at 
occupational profile $\bthetaM \in \BThetaM$, then $\Bz^*$ 
maximizes $\sum_{i \in \cali M} v_i(z_i,\thetaM_i)$ on 
$\BZ(\bthetaM)$, and thus is Pareto efficient at $\bthetaM$.  
\end{proposition}

\smallskip

Equilibrium prices need not be unique, and individual 
equilibrium payoffs depend on which one prevails.  
As it turns out, this issue affects existence of 
occupational equilibria in the market.  
For now, in order to provide a formal definition of what 
we mean by occupational equilibrium in a market economy, 
we follow \citet{Makowski:1995cf} in positing a 
\emph{price selection mapping}: a commonly known function 
$p \colon \BThetaM \to \R^L$ assigning, to each 
occupational profile $\bthetaM$, an equilibrium price that 
is correctly anticipated by all individuals when choosing 
their occupations.
Write $\Bz(\bthetaM) = (z_i(\bthetaM))$ for an associated 
equilibrium trade profile. 
Individual $i$'s \emph{market payoff} at $\bthetaM$ is then 
	\begin{equation}   \label{Eq:payoff-market}
		\begin{split}
		\bar\pi_i(\bthetaM) 
		& \ \equiv \ v_i(z_i(\bthetaM),\thetaM_i) \ - \ 
			p(\bthetaM) \cdot z_i(\bthetaM)   \\[.75ex] 
		& \ = \ \max_{z \in Z_i(\thetaM_i)} \big\{ v_i(z,
			\thetaM_i) \ - \ p(\bthetaM) \cdot z \big\} .
		\end{split}
	\end{equation}

An \emph{occupational equilibrium in the market} is then 
a profile of occupational choices that gives no individual
incentives to deviate unilaterally.

\begin{definition}  \label{D:market-occup-equi}
An \emph{occupational equilibrium in the market}
is a profile $\bsthetaM = (\sthetaM_i) \in \BThetaM$ such that, 
for each individual $i \in \cali M$ and each alternative 
occupation $\thetaM_i \in \ThetaM_i$,
	\begin{equation}  \label{Eq:market-OE}
		\bar\pi_i (\sthetaM_i,\bsthetaM_{-i})
		\ \ge \ \bar\pi_i (\thetaM_i,\bsthetaM_{-i}) .
	\end{equation}
\end{definition}

\subsection{Existence of Occupational Equilibria in the 
Market}  \label{S:occup-existence-market} 

An occupational equilibrium in the market is a Nash 
equilibrium of the game in which individuals choose occupations 
and anticipate the payoffs from a local market equilibrium, 
taking the price selection mapping as part of the structure 
of the game (and thus commonly known).  
Local market equilibria exist at every occupational profile, 
but this does not in itself establish existence of an 
occupational equilibrium, even when \autoref{A:structure} 
is imposed.  
A key in obtaining existence of occupational equilibria 
in the market is the possibility of securing a continuous 
price selection mapping. 

For each occupational profile $\bthetaM\in\BThetaM$, define 
the market economy \emph{equilibrium price correspondence}
$P \colon \BThetaM \rightrightarrows \R^L$ by 
	\begin{equation}  \label{Eq:price-correspondence}
		P (\bthetaM)
		\ \equiv \ \big\{ p \, \in \, \R^L \sth
			\text{$(\Bz,p)$ is a market equilibrium 
			at $\bthetaM$, for some $\Bz$} \big\} .
	\end{equation}
The next result is the market counterpart of 
\autoref{T:jungle-occup-pure}.  
As before, we assume that \autoref{A:structure}, as applied 
to $\cali M \ne \emptyset$, continues to hold.

\begin{theorem}   \label{T:market-occup-pure}  
Suppose there exists a continuous selection $p \colon \BThetaM 
\to \R^L$ of the equilibrium price correspondence $P$. 
\begin{enumerate}[leftmargin=1.5em,label=(\alph*)]
		\setcounter{enumi}{0} \itemsep.25ex
	\item For every $i \in \cali{M}$, the payoff function 
		$\bar\pi_i \colon \BThetaM \to \R$ in 
		\autoref{Eq:payoff-market} is continuous.
	\item If, in addition, for every individual $i \in \cali{M}$ 
		and every $\bthetaM_{-i} \in \times_{j \ne i} \ThetaM_j$, 
		the set of maximizers of $\thetaM_i \mapsto 
		\bar\pi_i (\thetaM_i,\bthetaM_{-i})$ over $\ThetaM_i$ 
		is convex, then a pure occupational equilibrium in 
		the market, in the sense of \autoref{D:market-occup-equi},
		exists.
\end{enumerate}

\end{theorem}

\begin{proof}
(a) From \autoref{A:structure}(b), for each $i \in \cali M$,  
the correspondence $Z_i \colon \ThetaM_i \rightrightarrows 
\R^L$ is continuous and has nonempty and compact values.  
\autoref{A:structure}(c) and the continuity of the price 
selection function $p$ imply that $i$'s objective function in 
\autoref{Eq:payoff-market}, understood as a parametric 
optimization problem, is finite and continuous on its graph.
Berge's Theorem of the Maximum therefore implies that 
$\bar\pi_i$ is continuous on $\BThetaM$.

\smallskip

(b) Define individual $i$'s best reply correspondence
	\begin{equation*}
		\BR_i(\bthetaM_{-i})
		\ \equiv \ \argmax \, \big\{ \bar\pi_i(\thetaM_i,
			\bthetaM_{-i}) \sth \thetaM_i 
			\in \ThetaM_i \big\} ,
		\end{equation*}
and define also the product correspondence
	\begin{equation*}
		\BR(\bthetaM)
		\ \equiv \ \times_{i \in \cali M} 
			\, \BR_i(\bthetaM_{-i}) .
	\end{equation*}
The arguments used in the proof of \autoref{T:jungle-occup-pure} 
apply here too, and thus Kakutani's Fixed Point Theorem 
provides an occupational profile $\bsthetaM \in \BThetaM$
such that $\bsthetaM \in \BR(\bsthetaM)$.  
This is a pure occupational equilibrium in the market economy. 
\end{proof}

Notice that our existence result in 
\autoref{T:market-occup-pure}(b) does not require strict 
concavity of individual valuations, strict convexity of 
individual feasible trade sets, or a continuous selection of
equilibrium trade profiles.  
At a given occupation and price, individual payoffs are the 
values of the corresponding maximization problems; 
continuity of these values follows directly from Berge's 
Theorem once a continuous price selection function is 
in place. 
Of course, this is an additional requirement.  
In particular, the selected equilibrium prices may change 
discontinuously with occupational choices even under 
\autoref{A:structure}.%
		\footnote{\citet[footnote~7]{Makowski:1995cf} point 
		out that concavity of individual valuations alone 
		does not guarantee existence of an occupational 
		equilibrium.}
On the other hand, our setting is sufficiently structured 
to guarantee some properties of the equilibrium price 
correspondence $P$, and points to ways of choosing a 
continuous selection. 
We turn our attention to these issues now.

\begin{proposition}   \label{P:price-correspondence} 
The equilibrium price correspondence $P \colon \BThetaM 
\rightrightarrows \R^L$ is upper hemicontinuous and has 
nonempty, compact and convex values that are contained in 
a common compact set.
\end{proposition}

\begin{proof}
Define $\mu \colon \R^L \times \BThetaM \to \R$ by 
	\begin{equation*}
		\mu (p,\bthetaM)
		\ \equiv \ \sum_{i \in \cali M} \max \, \Bigl\{ 
			v_i(z_i,\thetaM_i) \, - \, p \cdot z_i 
			\sth z_i \, \in \, Z_i(\thetaM_i) \Bigr\} .
		\end{equation*}
Berge's Theorem of the Maximum implies that $\mu$ is continuous.
Notice that $\mu$ is also convex in $p$ for fixed $\bthetaM$, 
since it is the sum of maxima of affine functions (in prices).

Fix an occupational profile $\bthetaM$ and a local market 
equilibrium $(\Bz^*,p^*)$.  
For any $p \in \R^L$, individual optimality and feasibility
of $\Bz^*$ show that 
	\begin{equation*}
		\mu (p,\bthetaM)
		\ \geq \ \sum_{i \in \cali M} v_i(z_i^*,\thetaM_i)
		\ = \ \mu (p^*,\bthetaM) . 
		\end{equation*} 
Thus, $p^*$ minimizes $\mu(\cdot,\bthetaM)$.  
Conversely, if $p' \in \R^L$ minimizes $\mu (\cdot,\bthetaM)$, 
then it must be that for all $i \in \cali M$, the trade 
position $z_i^*$ maximizes $v_i(z_i,\thetaM_i) - p' \cdot z_i$ 
on $Z_i(\thetaM_i)$. 
Otherwise, using feasibility of the trade profile $\Bz^*$, 
one has 
	\begin{align*}
		\mu (p^*,\bthetaM) 
		\ = \ \sum_{i \in \cali M} v_i(z_i^*,\thetaM_i) 
			\ - \ p' \cdot \sum_{i \in \cali M} z_i^* 
		\ < \ \mu (p',\bthetaM) ,
	\end{align*}
a contradiction.  
This shows that if $p'$ minimizes $\mu (\cdot,\bthetaM)$, then 
$(\Bz^*,p')$ is also a local market equilibrium. 
Consequently, we can write 
	\begin{equation*}
		P(\bthetaM)
		\ = \ \arg\min \big\{ \mu (p,\bthetaM) \sth 
			p \in \R^L \big\} . 
	\end{equation*} 
We conclude that the equilibrium price correspondence $P$ 
has values that are nonempty, by the existence of local 
market equilibria, and convex, by convexity of 
$\mu(\cdot,\bthetaM)$.

We next establish a uniform bound on equilibrium prices.  
Let $A(\bthetaM) = \sum_{i \in \cali M} Z_i(\thetaM_i)$.  
Seen as a correspondence defined on $\BThetaM$, $A$ 
is continuous and has compact, convex values.  
The compactness of $\BThetaM$ and \autoref{A:structure}(d)
imply that there is some $r > 0$ for which  
	\begin{equation*}
		\big\{ \bar z \, \in \, \R^L \sth \|\bar z\| \ \leq \ 
		r \big\} \, \subseteq \, A(\bthetaM),
			\qquad \text{for all $\bthetaM \in \BThetaM$.}
	\end{equation*}  
By \autoref{A:structure}(a) and (b), the graph of the 
individual trade correspondence $Z_i$ is compact, for all 
$i \in \cali M$.  
Since the market economy has finitely many individuals, 
\autoref{A:structure}(c) now furnishes a constant $K > 0$ 
such that
	\begin{equation*}
		\sum_{i \in \cali M} |v_i(z_i,\thetaM_i)|
		\ \leq \ K ,
	\end{equation*}
for every occupational profile $\bthetaM = (\thetaM_i)$ and 
every individually feasible trade profile $\Bz = (z_i)$.  
For $p = \zero$, note we have $\mu (\zero,\bthetaM) \leq K$.  
For $p \ne \zero$, choose individual trade positions that  
aggregate to $- r \, p/\|p\|$.  
With such a trade profile 
	\begin{equation*}
		\mu (p,\bthetaM) \ \geq \ r \,\|p\| \, - \, K . 
	\end{equation*}
Thus, for any occupational profile $\bthetaM$, a price 
$p \in \R^L$ minimizing $\mu(\cdot,\bthetaM)$ satisfies 
$\|p\| \leq 2K/r$.  
We may therefore restrict, without loss of generality, 
the minimization defining the equilibrium price correspondence 
$P$ to one common compact ball. 
Berge's Theorem of the Maximum gives the upper hemicontinuity 
and the compact value property for this correspondence.
\end{proof}

If $P$ is in addition lower hemicontinuous, 
then it is a continuous correspondence and thus 
admits a continuous selection \citep{Michael:1956}. 
Under such a selection, convex occupational best replies
as in \autoref{T:market-occup-pure}(b) ensure existence
of a pure occupational equilibrium.  
An alternative route to imposing this extra condition 
on the endogenous price correspondence is to add conditions 
on primitives so that equilibrium prices are unique.  
If there exists a unique equilibrium price vector 
for each occupational profile, the correspondence 
$P$ of \autoref{Eq:price-correspondence} is singleton-valued.  
\autoref{P:price-correspondence} now shows that we 
can treat it as a continuous equilibrium price function. 
Uniqueness can be obtained, for example, from the familiar 
supporting-price argument underlying the analysis of 
\citet{Makowski:1987rf}.  
It is enough that, at each occupational profile, some 
trade profile maximizing the sum of valuations gives an 
interior trade to an individual whose valuation  
is differentiable at that trade.

\begin{proposition}   \label{P:market-interior-trader}
Suppose that, for every occupational profile $\bthetaM \in 
\BThetaM$, there exist an individual $j \in \cali M$ 
and a trade profile 
	\begin{equation*}
		\hat{\Bz} \, \in \, \arg\max \, \Bigl\{
			\sum_{i \in \cali M} v_i(z_i,\thetaM_i) \sth 
			\Bz \, \in \, \BZ(\bthetaM) \Bigr\}  
	\end{equation*}
such that $\hat z_j \in \interior Z_j(\thetaM_j)$ and 
moreover $v_j(\cdot,\thetaM_j)$ is differentiable at 
$\hat z_j$. 
Then equilibrium prices are unique at every occupational 
profile, and all local market equilibria are payoff 
equivalent. 
\end{proposition} 

\begin{proof}
Fix $\bthetaM$ and choose $\hat{\Bz}$ and $j$ as in 
the statement of the proposition.  
For any $p \in P(\bthetaM)$, let $(\Bz,p)$ be any local market 
equilibrium.  
Individual optimal behavior gives 
	\begin{equation*}
		v_i(z_i,\thetaM_i) \, - \, p\cdot z_i
		\ \geq \ v_i(\hat z_i,\thetaM_i) \, - \, p \cdot 
			\hat z_i, 
			\qquad \text{for all $i \in \cali M$.}
		\end{equation*}
Summing these inequalities across individuals, and using 
the market clearing condition, the monetary terms 
on both sides of the aggregate expression cancel.  
Since $\hat{\Bz}$ maximizes the sum of valuations, 
the sums on both sides must be equal.  
Each individual inequality is therefore an equality, so 
$(\hat{\Bz},p)$ is also a local market equilibrium that 
generates the same individual equilibrium payoff to 
each $i \in \cali M$.  

Since $\hat z_j$ is an interior point of $Z_j(\thetaM_j)$,  
the differentiability of her valuation implies that 
	\begin{equation*}
		p \ = \ \nabla_z v_j(\hat z_j,\thetaM_j)
		\qquad  \text{for every $p \in P(\bthetaM)$.} 
	\end{equation*}
Since $\bthetaM$ was arbitrary, the equilibrium price 
correspondence is singleton-valued at every occupational 
profile. 
\end{proof}

Only one trade profile maximizing the sum of valuations 
and one individual need satisfy the condition at each 
occupational profile.  
Both may depend on $\bthetaM$.  
The individual's marginal valuation at her interior trade 
determines the equilibrium price, while continuity 
follows from upper hemicontinuity of $P$ and uniqueness. 
Neither the individual nor the maximizing trade profile
is required to be selected continuously.


\section{The Market and the Jungle Compared}   
	\label{S:comparison}

The results obtained so far identify the allocations and payoffs 
that individuals in the society anticipate when choosing 
occupations in the market and in the jungle.  
Suppose that \autoref{A:structure} holds and, in addition, 
individual valuations are strictly concave in trade positions, 
individual feasible trade sets are strictly convex, and 
the interiority and differentiability conditions in 
\autoref{P:market-interior-trader} are in place. 
In this case, the market has a unique local equilibrium at 
each occupational profile. 
Likewise, at each occupational profile, the jungle has a 
unique local equilibrium  generated by the serial 
dictatorship procedure for each hierarchy.  
Both outcomes determine, uniquely, individual market payoffs and 
individual expected jungle payoffs. 
Moreover, these payoff functions are continuous in their 
respective occupational profiles.  
When the sets of individual best replies are convex,
\autoref{T:jungle-occup-pure} and \autoref{T:market-occup-pure} 
show existence of pure occupational equilibria in both 
economies.%
		\footnote{Our existence results do not establish 
		uniqueness of occupational equilibria in either 
		the market or the jungle.}  

A main goal in our work is to consider the coexistence of the 
market and the jungle as a social equilibrium, where all 
individuals in our society decide which economy to join 
before making occupational choices.  
As a first pass, in this section we compare the welfare 
performance of the market and the jungle as two economic 
institutions operating on a level playing field.  
We first hold occupational choices fixed and use 
aggregate payoffs, i.e., the sum of individual payoffs 
generated by the market and by the jungle, respectively, 
as a yardstick.   
We next consider the performance of each economy at its 
occupational equilibria.

Throughout this section, we organize the entire society 
once as a market and once as a jungle. 
Namely, we assume $\cali M = \cali S = \cali J$, and further 
$\ThetaM_i = \Theta_i = \ThetaJ_i$ for every individual 
$i \in \cali S$.%
		\footnote{Except for this section, we treat $\cali M$ 
		and $\cali J$ as a partition of $\cali S$, albeit 
		leaving open the possibility 
		that one of these economies is empty.}  
This makes our comparisons here purely institutional: 
the two economies face the same population, same 
occupational opportunities, valuations and feasible trade sets.  
Any difference in performance is therefore due to the 
institutional rules, and not to occupational opportunities, 
technology, or preferences.    
An important institutional difference is that the market 
supports enforceable transfers of the numeraire, whereas 
the jungle allocates non-monetary goods through the 
hierarchy without allowing for monetary compensations.


\subsection{Profile by Profile Comparison}  
		\label{SS:profile-comparison}

Recall that the local equilibria in the jungle selected 
by the serial dictatorship procedure are Pareto 
efficient according to \autoref{D:local-efficiency}.  
Local market equilibria are also Pareto efficient in that 
sense; in addition, they satisfy the stronger property of 
maximizing the sum of individual valuations over feasible 
trade profiles (\autoref{P:local-m-efficiency}).  
The distinction, of course, comes from the fact that  
\autoref{D:local-efficiency} assumes monetary holdings among 
individuals fixed, whereas the market permits transfers 
of the numeraire (as long as the money market clears).

Since preferences are quasilinear in the numeraire, the 
sum of individual valuations provides a common money-metric 
measure of economic performance. 
For that purpose, it is convenient to define 
	\begin{equation}   \label{Eq:gains-from-trade}
		g ( \bthetaE ) 
		\ \equiv \ \max_{\Bz} \, \Bigl\{ \,\sum_{i \in \cali E} \,
			v_i(z_i,\thetaE_i) \sth \Bz \in \BZ(\bthetaE) \,\Bigr\}
	\end{equation}
as the \emph{maximal gains from trade} at occupational 
profile $\bthetaE$ in the economy $\cali E \in 
\{\cali M,\cali J\}$. 
Because the no-trade option is feasible for every 
occupational choice, this measure includes the values 
of no trade at the given occupations; the gains relative 
to no trade are $g(\bthetaE) - \sum_{i} v_i (\zero, 
\thetaE_i)$.
One readily observes that, under \autoref{A:structure}, 
for every $\bthetaE \in \BThetaE$ in a nonempty economy, 
maximal gains from trade in \autoref{Eq:gains-from-trade} 
are attained at some trade profile in $\BZ(\bthetaE)$.  
If, in addition, $v_i(\cdot,\thetaE_i)$ is strictly concave 
on $Z_i(\thetaE_i)$ for all $i \in \cali E$, then maximal 
gains from trade are attained at a unique trade profile.

We now state our first comparison of the performance of the 
jungle and the market. 
For this section, we write $\btheta = (\theta_i)$ for 
a profile in the common set of occupational profiles 
$\BTheta$, and use the superscripts $\cali M$ and $\cali J$ 
only to record which set of rules is applied in each 
circumstance.

\begin{proposition}   \label{P:market-beats-jungle}
Suppose that $\cali M = \cali S = \cali J$, and 
let $\ThetaM_i = \Theta_i = \ThetaJ_i$ for every individual 
$i \in \cali S$. 
Then, for every occupational profile $\bthetaM = \btheta 
= \bthetaJ$ in $\BTheta$, one has 
	\begin{equation*}
		\sum_{i \in \cali S} \bar\pi_i (\bthetaJ)
		\ \le \ \sum_{i \in \cali S} \bar\pi_i (\bthetaM) .
	\end{equation*}
\end{proposition}

\begin{proof} 
Starting with the jungle at occupational profile 
$\bthetaJ$, fix a hierarchy $h \in \scr H(\cali J)$ 
and let $\Bz \in \BZ^*(\bthetaJ,h)$ be any jungle equilibrium 
at $(\bthetaJ,h)$.  
By \autoref{C:local-j-properties}, we have $\pi_{h(n)} 
(\bthetaJ) = v_{h(n)}(z_{h(n)},\thetaJ_{h(n)})$ for 
every position $n = 1,\ldots,|\cali J|$ in this hierarchy. 
Since $\Bz \in \BZ(\bthetaJ)$, 
	\begin{equation*}
		\sum_{n = 1}^{|\cali J|} \, \pi_{h(n)} (\bthetaJ)
		\ \le \ g (\bthetaJ) .
	\end{equation*}
Averaging over all power relations to obtain individual jungle 
payoffs, as in \autoref{Eq:expec-payoff-jungle}, preserves 
the above inequality:
	\begin{equation}  \label{Eq:market-beats-jungle-1}
		\sum_{i \in \cali J} \bar\pi_i(\bthetaJ) 
		\ = \ \sum_{h \in \scr H(\cali J)} \frac{1}{|\cali J|!} 
			 \, \sum_{n = 1}^{|\cali J|} \pi_{h(n)} 
			(\bthetaJ)  
		\ \leq \ g (\bthetaJ) .     
	\end{equation}

Moving to the market at occupational profile 
$\bthetaM$, we know from \autoref{P:local-m-existence} and 
\autoref{P:local-m-efficiency}, that a local market 
equilibrium exists and its associated trade profile 
$\Bz(\bthetaM) = (z_i(\bthetaM))$ maximizes 
$\sum_i v_i(z_i',\thetaM_i)$ over $\BZ(\bthetaM)$.  
Thus we obtain $\sum_i v_i(z_i(\bthetaM),\thetaM_i) 
\, = \, g(\bthetaM)$.
The market clearing condition gives $\sum_i z_i(\bthetaM) 
= \zero$.
Summing individual payoffs, defined in \autoref{Eq:payoff-market}, 
across individuals in the market immediately yields  
	\begin{equation}  \label{Eq:market-beats-jungle-2}
		\sum_{i \in \cali M} \bar\pi_i (\bthetaM) 
		\ = \ \sum_{i \in \cali M} v_i(z_i(\bthetaM),\thetaM_i)
			\ - \ p(\bthetaM) \cdot \sum_{i \in \cali M}
			z_i(\bthetaM)
		\ = \ g (\bthetaM) . 
	\end{equation}

Recalling that $\cali M = \cali S =  \cali J$ and 
$\bthetaM = \btheta = \bthetaJ$, the result follows readily 
from combining \autoref{Eq:market-beats-jungle-1} and 
\autoref{Eq:market-beats-jungle-2}. 
\end{proof}

\autoref{Eq:market-beats-jungle-2} is the accounting 
identity of \cite{Makowski:1995cf} applied to our setting: 
market payoffs exhaust the maximal gains from trade at 
every occupational profile.  
In equilibrium, monetary payments cancel when individual 
payoffs are added, because the corresponding positions 
sum to zero.  
In the jungle, the hierarchical choice of trade positions 
can leave a smaller sum of valuations even though the 
selected outcome is Pareto efficient. 
This is what \autoref{P:market-beats-jungle} formalizes: 
at any given occupational profile, the same 
group of individuals are collectively weakly better off in the 
market than in the jungle. 
It does not imply that every individual prefers the market.

We push this comparison one more step and ask when 
it is the case that individuals are collectively strictly 
worse off in the jungle.  
The relevant tension is between the trade position 
that an individual would choose if constrained only by her own 
capacities, and those compatible with maximizing the 
sum of valuations.
To explore this, define individual $i$'s \emph{satiation set} 
at occupation $\theta_i \in \Theta_i$ by 
	\begin{equation}   \label{Eq:satiation}
		\Zsat_i (\theta_i)
		\ \equiv \ \arg\max 	\big\{ v_i(z_i,\theta_i) \sth 
			z_i \, \in \, Z_i(\theta_i) \big\}  .
	\end{equation}
This set contains all the trade positions $i$ would choose, 
given her occupation $\theta_i$, if she faced no constraints 
other than her own feasible trade set.  
Note that under our maintained assumptions, $\Zsat_i (\theta_i)$ 
is nonempty, compact and convex.

\begin{definition}   \label{D:plentiful-satiation}
Let $\bthetaE = (\thetaE_i) \in \BThetaE$ be an occupational 
profile for a nonempty economy $\cali E\in\{\cali M,\cali J\}$.
\begin{enumerate}[leftmargin=1.5em,label=(\alph*)]
		\setcounter{enumi}{0} \itemsep.25ex
	\item Resources are \emph{plentiful  at $\bthetaE$} if
		\begin{equation*}
			\zero \ \in \ 
			\sum_{i \in \cali E} \, \Zsat_i (\thetaE_i) .  
		\end{equation*}
    \item Resources allow \emph{the satiation of individual $i \in 
    	\cali E$ at $\bthetaE$} if
		\begin{equation*}
			\zero \ \in \ \Zsat_i (\thetaE_i) \ + \ 
				\sum_{j \ne i}  Z_j (\thetaE_j) .
		\end{equation*}
\end{enumerate}
\end{definition}

Both conditions in \autoref{D:plentiful-satiation} are 
stated in terms of individual feasible trade sets and 
valuations; neither refers to an equilibrium object.  
Resources are plentiful at $\bthetaE$ if there exists a trade 
profile containing a most-preferred trade position for 
each individual that can be feasibly provided by the economy.  
Thus, every individual in the economy can be simultaneously 
satiated.   
On the other hand, resources allow the satiation of an 
individual at $\bthetaE$ if there exists a feasible trade 
profile that gives said individual one of her most-preferred 
trades, without requiring that the others be satiated. 
When the latter condition holds for every individual, we 
say that resources allow the satiation of all individuals 
separately.

If resources are plentiful, the role played by either economic 
institution, the anonymous market or the non-anonymous jungle, 
is essentially the same: coordination of economic activities, 
without arbitration of wants by competitive prices or 
hierarchical choices.  
Indeed, the serial dictatorship procedure can satiate 
every individual under every hierarchy. 
Thus, if resources are plentiful at $\bthetaM = 
\btheta = \bthetaJ$, and $\cali M = \cali S = \cali J$, 
we have 
	\begin{equation*}
		\sum_{i \in \cali S} \bar\pi_i(\bthetaM) 
		\ = \ g (\btheta) \ = \ 
		\sum_{i \in \cali S} \bar\pi_i(\bthetaJ) .
	\end{equation*} 
In contrast, individual satiation need not be compatible 
with maximizing the sum of valuations.  
If an individual can be satiated in the jungle economy, 
but no trade profile maximizing aggregate surplus satiates 
her, placing her first in the hierarchy produces a strictly 
smaller aggregate surplus.  
Since every hierarchy has positive probability, one such 
individual is enough for the performance difference between 
the market and the jungle to be strict.

\begin{proposition}   \label{P:jungle-wedge}
For an occupational profile $\bthetaJ \in \BThetaJ$ in 
a nonempty jungle, the following hold.
\begin{enumerate}[leftmargin=1.5em,label=(\alph*)] 
		\setcounter{enumi}{0} \itemsep.25ex
    \item If $g(\bthetaJ) \, > \, \sum_{i\in\cali J} \bar\pi_i 
    	(\bthetaJ)$, then resources in the jungle are not 
    	plentiful at $\bthetaJ$.
    \item If resources allow the satiation of some 
    	$i \in \cali J$ at $\bthetaJ$, but no feasible 
    	trade profile attaining $g(\bthetaJ)$ contains  
    	a trade position for $i$ that belongs to 
    	$\Zsat_i(\thetaJ_i)$, then $g(\bthetaJ) \, > \, 
    	\sum_{j \in \cali J} \bar\pi_j (\bthetaJ)$.  
	\item In particular, the strict inequality holds 
		if $g(\bthetaJ)$ is attained at a unique trade 
		profile, resources are not plentiful, and resources 
		allow the satiation of all individuals separately.
\end{enumerate}
\end{proposition}

\begin{proof}
(a) We show the contrapositive. 
When resources are plentiful at $\bthetaJ$, there exists a 
feasible trade profile $\Bzsat = (\zsat_i) \in \BZ(\bthetaJ)$ 
such that $\zsat_i \in \Zsat_i(\thetaJ_i)$ for every 
$i \in \cali J$.
This profile maximizes each individual's valuation over  
her own feasible trade set, and thus 
	\begin{equation*}
	    g(\bthetaJ) \ = \ 
	    	\sum_{i\in\cali J}v_i(\zsat_i,\thetaJ_i).
	\end{equation*} 
	
Fix any hierarchy $h \in \scr H (\cali J)$.  
Since $\Bzsat$ belongs to $\BZ(\bthetaJ) = \BZ_{h(0)}(\bthetaJ)$ 
and maximizes the first dictator's valuation, it belongs to
$\BZ_{h(1)}(\bthetaJ)$.  
Recursively, if $\Bzsat$ belongs to $\BZ_{h(n-1)}(\bthetaJ)$, 
it also maximizes individual $h(n)$'s valuation over 
this constraint set, and hence belongs to $\BZ_{h(n)}
(\bthetaJ)$. 
This argument shows that $\Bzsat \in \BZ^*(\bthetaJ,h)$.
From \autoref{C:local-j-properties}, every local jungle 
equilibrium in $\BZ^*(\bthetaJ,h)$ gives each individual 
her satiation payoff.
This holds under every hierarchy, so averaging gives 
	\begin{equation*}
		\sum_{i \in \cali J} \bar\pi_i (\bthetaJ)
		\ = \ \sum_{i \in \cali J} v_i(\zsat_i,\thetaJ_i)
		\ = \ g(\bthetaJ) .
	\end{equation*}

\smallskip

(b) Suppose resources allow the satiation of $i \in \cali J$. 
Fix a hierarchy $h \in \scr H (\cali J)$ with $h(1) = i$, 
and let $\hat\Bz \in \BZ^*(\bthetaJ,h)$. 
Since $i$ is the most powerful individual, her maximal 
valuation over $Z_i(\thetaJ_i)$ can be attained by a 
feasible trade profile.  
Therefore, the first optimization problem in the recursion 
of \autoref{P:local-j-existence} with the hierarchy $h$ 
gives $\hat z_i \in \Zsat_i(\thetaJ_i)$.
Since no feasible trade profile attaining $g(\bthetaJ)$ 
includes $\hat z_i$, we have 
	\begin{equation*}
	    \sum_{j \in \cali J} v_j(\hat z_j,\thetaJ_j)
	    	\ < \ g(\bthetaJ).
	\end{equation*}
This holds under every hierarchy that places individual $i$ 
in the first position.  
Under every other hierarchy, the aggregate payoff is at 
most $g(\bthetaJ)$. 
Consequently,
	\begin{equation*}
	    \sum_{j \in \cali J} \bar\pi_j (\bthetaJ)
	    \ = \ \frac{1}{|\cali J|!} \, \sum_{h} \ 
	    	\sum_{n=1}^{|\cali J|} \pi_{h(n)}(\bthetaJ)
	    \ < \ g(\bthetaJ).
	\end{equation*}

\smallskip

(c) Let $\bar\Bz = (\bar z_j)$ be the unique trade profile 
attaining $g (\bthetaJ)$.  
Since resources are not plentiful at $\bthetaJ$, for some 
individual $i \in \cali J$ it must be true that $\bar z_i 
\notin \Zsat_i(\thetaJ_i)$.
Otherwise, $\bar\Bz$ would constitute a feasible trade 
profile satiating all individuals, which is not possible.  
On the other hand, resources allow the satiation of $i$ 
separately. 
Since $\bar\Bz$ uniquely attains $g (\bthetaJ)$, we conclude 
that no trade profile attaining $g (\bthetaJ)$ satiates her.  
The argument employed to prove (b) now applies.
\end{proof}

\autoref{P:jungle-wedge}(b) isolates what hierarchical, 
non-anonymous institutions cost in terms of welfare 
performance. 
Under individual satiation, there is an individual 
in the jungle who, when operating in the top position 
in the hierarchy, can coerce others to achieve her satiation 
point. 
But her objective is in direct conflict with the 
maximization of aggregate welfare, understood as the 
sum of valuations.  
The jungle has no mechanism to offer her compensation 
for choosing a trade profile that maximizes gains from 
trade instead of her satiation trade position. 
The serial dictatorship outcome remains Pareto efficient.  
Under the conditions of part (b), placing this individual 
first selects an allocation that does not maximize 
aggregate welfare.

The following sufficient condition makes the conflict 
identified in \autoref{P:jungle-wedge} explicit in terms 
of primitives: only one individual is required to have 
a unique interior satiation trade and a valuation that 
is differentiable there.  
The valuations of the other individuals need not
be strictly concave or differentiable.

\begin{corollary}  \label{C:jungle-wedge-prim}
Suppose resources in the jungle are not plentiful 
at occupational profile $\bthetaJ = (\thetaJ_i) \in 
\BThetaJ$. 
Suppose in addition that resources allow the satiation of 
an individual $i \in\cali J$ whose valuation $v_i(\cdot,
\thetaJ_i)$ admits a unique interior satiation trade, 
i.e., $\Zsat_i(\thetaJ_i) = \{\zsat_i\}$, with 
$\zsat_i \in \interior Z_i(\thetaJ_i)$.
If $v_i(\cdot,\thetaJ_i)$ is differentiable at $\zsat_i$, then
	\begin{equation*}
	    g(\bthetaJ) \ > \ \sum_{j \in \cali J} 
	    	\bar\pi_j(\bthetaJ).
	\end{equation*}
\end{corollary}

\begin{proof}
Let $\hat \Bz = (\hat z_j) \in \BZ(\bthetaJ)$ be any feasible trade 
profile satiating individual $i$, so $\hat z_i=\zsat_i$. 
Since resources are not plentiful, there exists some individual 
$j \ne i$ with $\hat z_j \notin \Zsat_j(\thetaJ_j)$. 
Choose any $\zsat_j \in \Zsat_j(\thetaJ_j)$ and let 
$\delta \, = \, v_j (\zsat_j,\thetaJ_j) \ - \ v_j 
(\hat z_j,\thetaJ_j) \, > \, 0$.

For $\epsilon >0$, define a trade profile $\Bz(\epsilon)$ by 
	\begin{equation*}
	    z_i(\epsilon) \ = \ \zsat_i \, - \, \epsilon \, 
	    	(\zsat_j - \hat z_j) ,
	    	\qquad 
	    z_j(\epsilon) \ = \ \hat z_j \, + \, \epsilon \,  
	    	(\zsat_j - \hat z_j) , 
	\end{equation*}
and $z_k(\epsilon) = \hat z_k$ for all $k \ne i,j$.
For sufficiently small $\epsilon >0$, this trade profile 
is feasible.
Individual $i$'s trade remains in her feasible set because
$\zsat_i$ is an interior point, individual $j$'s trade is
feasible by the convexity of $Z_j(\thetaJ_j)$, and the 
changes in their trades sum to zero.

Since $\zsat_i = \hat z_i$ is an interior maximizer and
$v_i(\cdot,\thetaJ_i)$ is differentiable at this trade 
position, we have $\nabla_z v_i(\zsat_i,\thetaJ_i) = \zero$, 
and thus we can write 
	\begin{equation*}
	    v_i(z_i(\epsilon),\thetaJ_i)
	    \, - \, v_i(\hat z_i,\thetaJ_i) \ = \ o(\epsilon).
	\end{equation*}
Concavity of $v_j(\cdot,\thetaJ_j)$ implies that 
	\begin{equation*}
	    v_j(z_j(\epsilon),\thetaJ_j) 
	    \, - \, v_j(\hat z_j,\thetaJ_j) \ \ge \ 
	    	\epsilon \, \delta .
	\end{equation*}
Hence, for sufficiently small $\epsilon > 0$,
	\begin{equation*}
		\sum_{k \in \cali J} v_k(z_k(\epsilon),\thetaJ_k)
		\ - \ \sum_{k \in \cali J} v_k(\hat z_k,\thetaJ_k)
		\ \ge \ \epsilon\,\delta \ + \ o(\epsilon) 
		\ > \ 0 .
	\end{equation*}
Hence no feasible trade profile satiating $i$ 
attains $g(\bthetaJ)$. 
\autoref{P:jungle-wedge}(b) now applies.
\end{proof}


\subsection{Equilibrium Comparison}  
		\label{SS:equilibrium-comparison}

\autoref{P:market-beats-jungle}, and the strict 
version that follows either \autoref{P:jungle-wedge} 
or \autoref{C:jungle-wedge-prim},  compare the welfare 
performance of the market and the jungle economies 
profile by profile, not at their respective 
equilibria.  
But individuals optimally choose their occupations in 
each economy, anticipating the optimizing behavior 
of others in each context.  
And the rules that regulate economic interactions 
for the market and the jungle are different.  
In the market, each individual best responds to the 
market payoffs given in \autoref{Eq:payoff-market}, which 
value an occupation at what it earns net of prices.  
In the jungle, each individual best responds to the expected 
serial dictatorship payoff in \autoref{Eq:expec-payoff-jungle}, 
which values an occupation at what its holder can take 
given her position in the hierarchy, averaged over all  
the positions she may draw.  
So, in general, the occupational equilibria of the two 
economies differ, even in the common environment we 
analyze in this section, where the economies are populated 
by the same individuals, and individuals have the 
same occupational opportunities in either of them.

Our main result in this section states that when the 
market economy is perfectly competitive, in a sense that 
we make precise below, it outperforms the jungle economy 
when both institutions operate at equilibrium. 
As readers familiar with their work will surely recognize, 
to obtain our institutional comparison at equilibrium 
positions we adapt ideas about the meaning of perfect 
competition in \cite{Makowski:1995cf} and 
\cite{Makowski:1987rf}.  
In the next definition, we take $p \colon \BThetaM \to \R^L$ 
to be a continuous selection of the equilibrium price 
correspondence $P$ defined in \autoref{Eq:price-correspondence}. 
Local market equilibria are those associated with this 
selection.

\begin{definition}  \label{D:perfect-competition}
An occupational equilibrium in the market $\bsthetaM = 
(\sthetaM_j) \in \BThetaM$ is \emph{perfectly competitive} 
if there exists a price vector $p^* \in \R^L$ such that 
	\begin{equation*}
		p^* \ = \ p(\thetaM_i,\bsthetaM_{-i}) ,
			\qquad \text{for all $\thetaM_i \in \ThetaM_i$, 
				all $i \in \cali M$.}
	\end{equation*}
\end{definition}

If an occupational equilibrium in the market is perfectly 
competitive, no individual can influence the equilibrium 
prices selected by $p$ through an uncoordinated 
change in her occupational choices.  
Thus, under perfect competition, every individual 
acts as a \emph{price-taker}, given others' equilibrium 
occupational choices, and cannot affect prices either 
by changing what she trades or by changing what she 
selects as occupation.

\begin{theorem}   \label{T:comparison-equilibrium}
Suppose that $\cali M = \cali S = \cali J$ and 
$\ThetaM_i = \Theta_i = \ThetaJ_i$ for every individual 
$i \in \cali S$. 
Let $\bsthetaM = (\sthetaM_i)$ be any occupational equilibrium 
for the market and $\bsthetaJ = (\sthetaJ_i)$ be any 
occupational equilibrium for the jungle. 
If $\bsthetaM$ is perfectly competitive, then 
	\begin{equation*}
		\sum_{i \in \cali S} \bar\pi_i (\bsthetaJ)
		\ \le \ \sum_{i \in \cali S} \bar\pi_i (\bsthetaM) .
	\end{equation*}
\end{theorem}

\begin{proof} 
Starting with the market economy, let $p^*$ be the price 
vector associated with the perfectly competitive 
occupational equilibrium $\bsthetaM$.  
Now take any occupational profile $\bthetaM \in \BThetaM$ 
and choose $\bar\Bz = (\bar z_j) \in \BZ(\bthetaM)$ 
attaining $g(\bthetaM)$. 
For every $i \in \cali M$, since $\bar z_i \in Z_i(\thetaM_i)$,
individual optimal behavior in the market at 
$(\thetaM_i,\bsthetaM_{-i})$ with equilibrium prices $p^*$ 
implies 
	\begin{equation*}
		\bar\pi_i (\thetaM_i,\bsthetaM_{-i})
		\ \geq \ v_i(\bar z_i,\thetaM_i) \, - \, 
			p^* \cdot \bar z_i .  
	\end{equation*}
Adding the above inequality across all individuals in 
$\cali M$, using $\sum_i \bar z_i = \zero$, and subtracting 
$\sum_i \bar\pi_i (\bsthetaM) = g(\bsthetaM)$ from both sides 
of the resulting expression results in  
	\begin{align*}
		0 \ \geq \ 
		\sum_{i \in \cali M} \Bigl[ \bar\pi_i (\thetaM_i,
			\bsthetaM_{-i}) \ - \ \bar\pi_i (\bsthetaM) \Bigr]
		\ \geq \ g(\bthetaM) \ - \ g(\bsthetaM) ,
		\end{align*}
where the first inequality follows from the fact that 
$\bsthetaM$ is an occupational equilibrium in the market. 
Since $\bthetaM$ was an arbitrary occupational profile, 
we conclude that $\bsthetaM$ maximizes gains from trade 
over all occupational profiles.

Going to the jungle, every local jungle equilibrium 
allocation at $\bsthetaJ$ is feasible and thus 
has aggregate surplus bounded above by $g (\bsthetaJ)$. 
Since the two institutions face the same occupational 
opportunities and the same population, i.e., $\BThetaM 
= \BTheta = \BThetaJ$, averaging over hierarchies and 
applying the preceding conclusion yields 
	\begin{equation*}
		\sum_{i\in\cali S} \bar\pi_i(\bsthetaJ)
		\ \leq \ g(\bsthetaJ) 
		\ \leq \ g(\bsthetaM)
		\ = \ \sum_{i \in \cali S} \bar\pi_i(\bsthetaM) .
			\qedhere
	\end{equation*}
\end{proof}

Of course, under the additional conditions of 
\autoref{P:jungle-wedge} or \autoref{C:jungle-wedge-prim}, 
the conclusion of \autoref{T:comparison-equilibrium} 
can be strengthened to show that aggregate surplus in a 
perfectly competitive occupational equilibrium in the 
market is strictly higher than aggregate surplus in the 
corresponding equilibrium in the jungle.

To gain intuition for our comparison results, let 
	\begin{equation*}  
		\Delta (\bthetaJ) 
		\ \equiv \ g(\bthetaJ) \ - \ \sum_{i \in \cali J} 
			\bar\pi_i (\bthetaJ) \ \ge \ 0 
			\andeq
		g^* \ \equiv \ \max_{\btheta \in \BTheta} 
			\, g (\btheta) .
	\end{equation*}
The first expression can be understood as 
the jungle's \emph{trade wedge} at $\bthetaJ$. 
It is the difference between maximal gains from trade 
at $\bthetaJ$ and the expected aggregate payoffs
generated by the serial dictatorship. 
It is always non-negative by \autoref{P:market-beats-jungle}, 
and strictly positive under the circumstances described 
in \autoref{P:jungle-wedge} and \autoref{C:jungle-wedge-prim}.    
The term $g^*$ captures the largest maximal gains from 
trade that can be generated by an occupational profile in 
$\BTheta$.

Let $\BThetaM = \BTheta = \BThetaJ$ again.  
Adding and subtracting $g^*$ in the expression  
$\sum_i \bar\pi_i(\bsthetaM) \, - \, \sum_i \bar\pi_i(\bsthetaJ)$, 
using \autoref{Eq:market-beats-jungle-2}, and rearranging 
gives 
	\begin{equation}   \label{Eq:wedge-decomposition}
		\sum_{i \in \cali S} \bar\pi_i (\bsthetaM) 
			\ - \ \sum_{i \in \cali S} \bar\pi_i (\bsthetaJ)
		\ = \ \bigl[ g^* \, - \, g(\bsthetaJ) \bigr]
			\ - \ \bigl[ g^* \, - \, g(\bsthetaM) \bigr]
			\ + \ \Delta(\bsthetaJ) . \ 
	\end{equation}

This identity separates the difference in institutional 
performance into three components.  
The first two components respond to equilibrium occupational 
choices in the jungle and in the market, respectively.    
Indeed, the \emph{occupational wedge} $g^* - g(\bsthetaE)$ 
measures the gap between the largest gains from trade that 
can be attained in this society, and those realized at the 
occupational equilibrium for the economy $\cali E \in 
\{\cali M, \cali J\}$.  
The occupational wedge is by definition non-negative, for any 
occupational equilibrium in either economy.  
It captures the failures of each economy to implement, in  
equilibrium, the occupations that would generate $g^*$.  
The third component is the local trade wedge in the 
jungle at its occupational equilibrium.  
This is what \autoref{Eq:market-beats-jungle-1} measures: at 
$\bsthetaJ$, each selected trade profile is Pareto
efficient, but need not maximize the sum of valuations.

The decomposition in \autoref{Eq:wedge-decomposition} 
settles what our profile by profile comparison in 
\autoref{SS:profile-comparison} does, and does not, imply.  
Even when at any given occupational profile the market 
performs better than the jungle, as 
\autoref{P:market-beats-jungle} establishes, it is still possible 
that at their respective equilibrium occupations, 
the jungle is collectively better off than the market. 
Indeed, this would happen if the market occupational 
wedge $g^* - g(\bsthetaM)$ is sufficiently large.
The pointwise comparison of \autoref{P:market-beats-jungle} 
is compatible with either outcome at equilibrium: the 
jungle dominates precisely when the market's occupational 
wedge exceeds the jungle's own occupational wedge together 
with its local trade wedge.  
The decomposition in \autoref{Eq:wedge-decomposition} also
pins down the effects of perfect competition in the market.  
If the occupational equilibrium $\bsthetaM$ is perfectly 
competitive, and all potential occupational opportunities 
in the jungle are also present in the market, 
then the market occupational wedge  
vanishes, and the difference between market performance 
and jungle performance at equilibrium unambiguously 
favors the market.  
This is the message of \autoref{T:comparison-equilibrium}.

Both comparisons concern aggregate payoffs: at a 
common occupational profile, and at occupational 
equilibria when the market equilibrium is perfectly 
competitive, aggregate market payoffs are at least as 
high as aggregate jungle payoffs.  
It is still possible, of course, for some individuals to 
be better off (even strictly) in the jungle economy. 
These individual comparisons are relevant for a future 
extension of our work, where we allow the market and 
the jungle to coexist with endogenous sizes and 
composition.

Nonetheless, our level-playing-field setting in this 
section permits individual comparison in a special case 
that is of independent interest.  
From \autoref{A:structure}(b), the null trade 
position $\zero$ is individually feasible for any occupational 
choice $\theta_i \in \Theta_i$.  
Therefore, when occupational choices in the market and 
in the jungle are the same as in the whole society, at 
any market occupational equilibrium $\bsthetaM$ we have 
	\begin{equation}  \label{Eq:market-outside-opt}
		\bar\pi_i (\bsthetaM) 
		\ \geq \ \bar\pi_i (\theta_i,\bsthetaM_{-i}) 
		\ \geq \ v_i(\zero,\theta_i)	 
	\end{equation}
for any possible deviation $\theta_i \in \Theta_i$, since the 
null trade position is affordable for $i$ at whatever equilibrium 
prices prevail in the market following her deviation 
to $\theta_i$.%
		\footnote{Thus, here we are not assuming that the 
		occupational equilibrium $\bsthetaM$ is perfectly 
		competitive.} 

Recall also that individual payoffs in the jungle are 
obtained by taking expectations over the family 
of all possible hierarchies in $\cali J$.  
Given $\bsthetaJ$, let $z_i(\bsthetaJ,h)$ denote the 
trade position for individual $i$ selected by the 
hierarchy $h \in \scr H(\cali J)$ as part of its local 
jungle equilibrium allocation, and let $z_i^e(\bsthetaJ)$ 
denote its expectation, i.e., $z_i^e(\bsthetaJ) \, \equiv \,  
\E_{\scr H (\cali J)} \big[ z_i(\bsthetaJ,h) \big]$.

\begin{proposition}  \label{P:comparison-special-case}
Suppose that $\cali M = \cali S = \cali J$ and 
$\ThetaM_i = \Theta_i = \ThetaJ_i$ for every individual 
$i \in \cali S$. 
Let $\bsthetaM = (\sthetaM_i)$ and $\bsthetaJ = (\sthetaJ_i)$ 
be occupational equilibria in the market and the jungle, 
respectively. 
\begin{enumerate}[leftmargin=1.5em,label=(\alph*)] 
		\setcounter{enumi}{0} \itemsep.25ex
	\item If individual $i$'s expected equilibrium allocation 
		in the jungle is $z_i^e(\bsthetaJ) = \zero$, then 
		she is weakly better off in the market than in the 
		jungle: $\bar\pi_i (\bsthetaJ) \, \leq \, 
		\bar\pi_i (\bsthetaM)$. 
	\item If all individuals in the society are identical, 
		in terms of occupational choices, preferences, and 
		individual feasible trade sets, and if the 
		occupational equilibrium $\bsthetaJ$ is symmetric, 
		then every individual is weakly better off 
		in the market than in the jungle. 
\end{enumerate}

\end{proposition}

\begin{proof}
(a) Suppose that $z^e_i(\bsthetaJ) = \zero$. 
Since $v_i(\cdot,\sthetaJ_i)$ is a concave function, 
it follows from the definition of the expected jungle payoff 
function in \autoref{Eq:expec-payoff-jungle} that   
	\begin{align*}
		\bar\pi_i (\bsthetaJ) 
		& \ = \ \E_{\scr H(\cali J)} \bigl[ \,
			v_i(z_i(\bsthetaJ,h),\sthetaJ_i) \, \bigr] 
				\\[.75ex]
		& \ \leq \ v_i(z_i^e(\bsthetaJ),\sthetaJ_i)  \\[.75ex]
		& \ = \ v_i(\zero,\sthetaJ_i) 
		  \ \leq \ \max \big\{ v_i(\zero,\theta_i) \sth 
			\theta_i \in \Theta_i \big\} 
		  \ \leq \ \bar\pi_i (\bsthetaM) ,
	\end{align*}
where the last inequality follows from the observation 
we made in \autoref{Eq:market-outside-opt}.

\smallskip

(b) Suppose that $\bsthetaJ$ is a symmetric occupational 
equilibrium in the jungle.  
Each individual then selects the same occupation, has the 
same feasible trade set, and has the same valuation 
function.
Relabeling individuals leaves the serial dictatorship 
problems unchanged.  
Consequently, the payoff associated with each position
in the hierarchy does not depend on the identity of the 
individual occupying it. 
Uniform randomization over hierarchies therefore gives
every individual the same expected payoff.
Indeed, for any hierarchy $h \in \scr H(\cali J)$, concavity of 
the common valuation function and market clearing imply, 
for every individual $i \in \cali J$,
	\begin{equation*}
		\tfrac{1}{|\cali J|} \sum_{j \in \cali J}
			v_j \bigl( z_j(\bsthetaJ,h),\sthetaJ_j \bigr)
		\ \leq \ v_i \Bigl( 	\tfrac{1}{|\cali J|}
			\sum_{j \in \cali J} z_j(\bsthetaJ,h),\sthetaJ_i
			\Bigr)
		\ = \ v_i(\zero,\sthetaJ_i).
		\end{equation*}
Taking expectations, the expression on the left equals 
the average expected jungle payoff and hence, by symmetry, 
each individual's expected jungle payoff. 
Thus $\bar\pi_i (\bsthetaJ) \leq v_i(\zero,\sthetaJ_i)$.
\autoref{Eq:market-outside-opt} now gives $\bar\pi_i 
(\bsthetaJ) \, \leq \, \bar\pi_i (\bsthetaM)$ 
for each $i$ in the society. 
\end{proof}

A few comments regarding \autoref{P:comparison-special-case} 
are worth mentioning.  
First, suppose valuations are strictly concave.  
Then the comparison in part (a) is strict whenever  
individual $i$'s equilibrium trades in the jungle vary 
across hierarchies.  
Under the symmetry conditions in part (b), the comparison 
is strict for every individual whenever the equilibrium 
jungle trade lottery is non-degenerate.  
Strict concavity therefore gives a strict welfare 
comparison when the hierarchy generates uncertainty over 
individual trades.  
Under these conditions, when the society is populated by 
identical individuals, each of them is worse off (and 
thus also collectively worse off) in a symmetric 
occupational equilibrium in the jungle than in any 
occupational equilibrium in the market.  
Second, this does not assume that the market occupational 
equilibrium is perfectly competitive. 
A perfectly competitive occupational equilibrium 
in the market realizes the largest gains from trade.  
Instead, what goes on in \autoref{P:comparison-special-case} 
is that the symmetric occupational equilibrium 
in the jungle underperforms any occupational equilibrium 
in the market.

Third, note that with identical individuals and concave 
valuations, a symmetric occupational equilibrium in the 
market can be implemented with zero trade at the selected 
prices.  
In this case, the market as an institution performs 
at least as well as the jungle not because it generates 
gains from trade, 
but because it respects the rights of the individuals to 
choose their null trade positions.  
Finally, in our model, the fact that identical individuals 
populate the society does not mean that the only possible 
occupational equilibria are symmetric. 
Different occupational choices may generate different 
marginal valuations of the commodities, and thus both the 
jungle economy and the market economy can sustain 
asymmetric equilibria (with ex post differentiation). 
In particular, the market can support beneficial exchange 
between individuals who were identical before making 
occupational decisions.

Some of these points are illustrated in the next example.

\begin{example}  \label{Ex:identical-individuals}
There are two individuals and two non-monetary commodities. 
Both individuals, $i = 1,2$, have the same set of 
occupations $\Theta_i=[-1,1]$, the same feasible trade 
correspondence $Z_i \colon \Theta_i \rightrightarrows \R^2$ 
given by 
	\begin{equation*}
		Z_i(\theta_i)
		\ = \ \{z_i \, \in \, \R^2 \sth 
			z_{i1}^2 \, + \, z_{i2}^2 \, \leq \, 9\},
	\end{equation*}
and the same valuation function $v_i \colon \R^2 \times 
\Theta_i \to \R \cup \{-\infty\}$, which assigns the value 
	\begin{equation*}
		v_i(z_i,\theta_i)
		\ = \ 3 \, + \, (1 + \theta_i)z_{i1} \, + \, 
			(1-\theta_i)z_{i2} \, - \, \tfrac12\big(z_{i1}^2
			\, + \, z_{i2}^2+\theta_i^2\big) 
	\end{equation*}
to any trade that belongs to $Z_i(\theta_i)$ and 
$- \infty$ otherwise.

Notice that a positive occupation raises the marginal valuation 
of the first good and lowers that of the second; 
a negative occupation reverses these effects.
The term $\theta_i^2/2$ can be interpreted as the cost of 
the occupational choice. 
These two individuals are identical. 
Moreover, their feasible trade sets remain identical at every
occupational profile, although, and this detail is 
important, different occupational choices induce different
marginal values over trades. 
All the conditions in \autoref{A:structure} are met; in 
addition, valuations are strictly concave and
differentiable in trade and the feasible trade sets are strictly 
convex.

Starting with the market economy, at an occupational profile 
$\btheta = (\theta_1,\theta_2)$, the unique market 
equilibrium consists of prices 
	\begin{equation*}
		p(\btheta)
		\ = \ \left(1 \, + \, \tfrac{1}{2}(\theta_1 \, + \, 
			\theta_2), \, 1 \, - \, \tfrac{1}{2} 
			(\theta_1 \, + \, \theta_2) \right)
		\end{equation*}
and trades
	\begin{equation*}
		z_i(\btheta) 
		\ = \ \left(\tfrac{1}{2}(\theta_i \, - \, \theta_j),
	    	\, - \, \tfrac{1}{2}(\theta_i \, - \, \theta_j) 
	    	\right),
			\qquad \text{where $j \ne i$.} 
	\end{equation*}
These trades are in the interior of the feasible sets, 
maximize each individual's valuation net of prices, and clear 
the two commodity markets. 
Individual $i$'s market payoff is then 
	\begin{equation*} 
		\bar\pi_i(\bthetaM) 
		\ = \ 3 \ + \ \tfrac{1}{4} (\theta_i-\theta_j)^2 
			\ - \ \tfrac{1}{2} \theta_i^2 .
		\end{equation*}

Individual $i$'s market payoff function is strictly concave 
in $\theta_i$. 
It is readily seen that, for each $\theta_j \in \Theta_j$, 
its unique maximizer is $\theta_i = -\theta_j$. 
Hence the market occupational equilibria are all profiles 
$\bthetaM = (\theta, - \theta)$, with $\theta \in [-1,1]$. 
At any such profile, each individual obtains a market 
payoff equal to $3 + \theta^2/2$.  
In particular, $\bsthetaM = (1,-1)$ is a market occupational 
equilibrium with 
	\begin{equation*}
		z_1(\bsthetaM) \ = \ (1,-1), \qquad 
		z_2(\bsthetaM) \ = \ (-1,1), \andeq
		p(\bsthetaM) \ = \ (1,1) . 
	\end{equation*}
Each individual supplies one unit of one good and receives 
one unit of the other, and both obtain a market payoff 
equal to $7/2$. 
In contrast, their best market payoff without trade is $3$, 
which is attained at the occupational profile 
$\bcircthetaM = (0,0)$. 
In this case, the individuals are identical after 
selecting their occupations and the market equilibrium involves 
no trade.   
Hence, the occupational equilibrium $\bsthetaM = (1,-1)$ 
generates a strict gain over the best zero-trade outcome 
for each individual, because it allows them to realize 
gains from trade after specialization.

Going to the jungle now, at any occupational profile 
$\btheta = (\theta_1,\theta_2)$, if individual $i$ is first in 
the hierarchy, she selects her satiation trade 
	\begin{equation*}
		\zsat_i(\btheta) 
		\ = \ (1 \, + \, \theta_i, \, 1 \, - \, \theta_i) .
	\end{equation*} 
The other individual is coerced into taking $- \zsat_i(\btheta)$ 
as her trade.  
These trades are individually feasible and clear the jungle. 
Averaging over the two possible hierarchies gives individual 
$i$'s jungle expected payoff: 
	\begin{equation*}
		\bar\pi_i(\bthetaJ)
		\ = \ 2 \ - \ \theta_i\,\theta_j \ - \ 
			\tfrac{1}{2} \theta_j^2 .
	\end{equation*}
For fixed $\theta_j$, individual $i$'s jungle payoff is 
linear in $\theta_i$.  
Her best reply correspondence takes the single value $-1$ 
when $\theta_j > 0$, $1$ when $\theta_j < 0$, and the entire 
interval $[-1,1]$ when $\theta_j = 0$. 
Readily, all best-reply sets are convex, as they are in the
market.

The profile $\bsthetaJ = (0,0)$ is a symmetric jungle 
occupational equilibrium. 
At this equilibrium, individual $i$ obtains trade $(1,1)$ 
and payoff $4$ when she is the first in the hierarchy, and 
trade $(-1,-1)$ and payoff $0$ when she is the second.  
Consequently,
	\begin{equation*}
		z_i^e (\bsthetaJ) \ = \ \zero  
			\andeq
		\bar\pi_i(\bsthetaJ) \ = \ 2 ,
			\qquad \text{for $i = 1,2$.}
	\end{equation*}
Comparing $\bsthetaJ$ with $\bsthetaM$ and $\bcircthetaM$ 
shows that 
	\begin{equation*}
		\bar\pi_i (\bsthetaJ) \ = \ 2 \ < \ 3 \ = \ 
		\bar\pi_i (\bcircthetaM) \ < \ \tfrac{7}{2} 
		\ = \ \bar\pi_i(\bsthetaM),
		\qquad i=1,2.
	\end{equation*}
The first inequality captures the loss from random 
coercive trade in the jungle equilibrium $\bsthetaJ$ 
relative to the zero-trade market equilibrium $\bcircthetaM$. 
Note that this is purely driven by institutional rules, 
as in these two equilibria individuals choose the same 
occupations and are thus ex post identical. 
The second inequality captures the gain supported by 
occupational differentiation and voluntary exchange 
in the market equilibrium $\bsthetaM$.

There are two other occupational equilibria in this jungle: 
$(1,-1)$ and $(-1,1)$, each giving expected payoff $5/2$ 
to both individuals.  
Thus, in this example, every market occupational equilibrium 
is strictly preferred to every jungle occupational 
equilibrium by each individual: welfare comparison does not 
depend on selecting the symmetric jungle equilibrium.  
\hfill $\diamond$
\end{example}


\bibliography{market-jungle-references}


\appendix

\begin{center}
	\bfseries \LARGE Appendix
\end{center}


\section{Payoff Dominance of Local Jungle Equilibria}
		\label{S:app-payoff-dom}

In this appendix we show that, under an additional 
assumption, the local jungle equilibria identified in 
\autoref{P:local-j-existence} are weakly payoff dominant. 
Fix a power relation $h \in \scr H (\cali J)$ for a 
nonempty jungle. 
Recalling that $\BZ_{h(0)}(\bthetaJ) \equiv \BZ(\bthetaJ)$, 
define
	\begin{equation*}
	    \BY_{h(n)}(\bthetaJ)
	    \ \equiv \ \BZ_{h(n-1)}(\bthetaJ) \setminus 
	    	\BZ_{h(n)}(\bthetaJ), 
	    	\qquad \text{for all $n = 1,\ldots,|\cali{J}|$.} 
	\end{equation*}
These are the trade profiles discarded at step $n$ of the
recursion in the proof of \autoref{P:local-j-existence}:
they give every stronger individual her payoff 
in the serial dictatorship, but provide individual $h(n)$ 
strictly less than $\pi_{h(n)}(\bthetaJ)$.

\begin{assumption}  	\label{A:hierarchical-improvement}
An occupational profile $\bthetaJ \in \BThetaJ$ satisfies 
improvement along the hierarchy $h \in \scr H (\cali J)$ 
if, for every $m \in \{1,\ldots,|\cali{J}| - 1\}$ and every 
$\hat{\Bz} \in \BY_{h(m)}(\bthetaJ)$, there exist $n \in 
\{m + 1, \ldots, |\cali{J}|\}$ and a feasible trade profile
$\tilde{\Bz}\in\BZ(\bthetaJ)$ such that 	
$v_i (\tilde z_i,\thetaJ_i) \, = \, v_i (\hat z_i,\thetaJ_i)$  
for all $i \in \cali J$, $i \ne 	h(m),h(n)$, and 
$v_{h(m)} \bigl(\tilde z_{h(m)},\thetaJ_{h(m)}\bigr) 
\, > \, v_{h(m)} \bigl(\hat z_{h(m)}, \thetaJ_{h(m)} \bigr)$.
\end{assumption}

The above condition requires an improving deviation only at
profiles discarded before the final step of the recursion,
and only against some weaker individual. 
The trade positions of third parties may change, provided 
their payoffs remain unchanged, as permitted by
\autoref{D:local-j-equilibrium}.    
Under improvements along the hierarchy $h$, the 
local equilibria in $\BZ^*(\bthetaJ,h)$ payoff dominate 
any other jungle equilibrium at $\bthetaJ$.

\begin{proposition}	\label{P:payoff-dom}
Suppose \autoref{A:structure} holds and fix a power 
relation $h \in \scr H (\cali J)$. 
Suppose the occupational profile $\bthetaJ \in \BThetaJ$ 
satisfies \autoref{A:hierarchical-improvement} relative 
to $h$.
\begin{enumerate}[leftmargin=1.5em,label=(\alph*)] 
		\setcounter{enumi}{0} \itemsep.25ex
	\item If $\hat{\Bz}$ is a jungle equilibrium for 
		$(\bthetaJ,h)$, then for all $n = 1,\ldots,|\cali{J}|$,
		\begin{equation}   \label{Eq:payff-dom}
		    v_{h(n)} \bigl(\hat z_{h(n)},\thetaJ_{h(n)}\bigr)
		    \ \le \ \pi_{h(n)}(\bthetaJ) . 
		\end{equation}
	\item \autoref{Eq:payff-dom} holds with equality for 
		every position $n < |\cali{J}|$ in the hierarchy. 
		For the weakest individual, the inequality is strict 
		if and only if $\hat{\Bz} \not\in \BZ^*(\bthetaJ,h)$.
	\item If, in addition, $v_i(\cdot,\thetaJ_i)$ is strictly 
		concave on $Z_i(\thetaJ_i)$ for all $i \in \cali J$, 
		then there is a unique local jungle equilibrium 
		at $\bthetaJ$. 
\end{enumerate}
\end{proposition}

\begin{proof}
(a)--(b) We first claim that $\hat{\Bz} \in \BZ_{h(|\cali{J}|-1)} 
(\bthetaJ)$, the constraint set of the weakest individual in 
the serial dictatorship recursion.  
Indeed, since $\hat{\Bz}$ is a jungle equilibrium, it belongs 
to $\BZ(\bthetaJ)$. 
By the nested construction of the solution sets, we have 
	\begin{equation*}
	    \BZ(\bthetaJ) \setminus
		    \BZ_{h(|\cali{J}|-1)} (\bthetaJ)
	    \ = \ \bigcup_{n = 1}^{|\cali{J}|-1} \BY_{h(n)}(\bthetaJ).
	\end{equation*}
If $\hat{\Bz}$ were not in $\BZ_{h(|\cali{J}|-1)}
(\bthetaJ)$, it would belong to $\BY_{h(m)}(\bthetaJ)$ 
for some $m < |\cali{J}|$.  
From \autoref{A:hierarchical-improvement}, there is 
a weaker individual $h(n)$, with $n > m$, and a feasible 
trade profile $\tilde{\Bz}$ that strictly increases the 
payoff of individual $h(m)$ while leaving every individual 
other than $h(m)$ and $h(n)$ at her payoff under $\hat{\Bz}$. 
This contradicts $\hat{\Bz}$ being a local jungle 
equilibrium.  

Since $\hat{\Bz} \in \BZ_{h(|\cali{J}|-1)} (\bthetaJ)$, 
we have
	\begin{equation*}
	    v_{h(n)} \bigl(\hat z_{h(n)},\thetaJ_{h(n)}\bigr)
	    \ = \ \pi_{h(n)}(\bthetaJ) ,
	    	\qquad \text{for all $n = 1,\ldots,|\cali{J}|-1$.}
	\end{equation*}
The weakest individual's payoff from the serial dictatorship 
is the maximum of her payoff over the same set:
	\begin{equation*}
	    \pi_{h(|\cali{J}|)}(\bthetaJ) 
	    \ = \ \max \,\Bigl\{ v_{h(|\cali{J}|)} \bigl(
	        z_{h(|\cali{J}|)}, \thetaJ_{h(|\cali{J}|)} \bigr)
	        \sth \Bz \in \BZ_{h(|\cali{J}|-1)} 
	        (\bthetaJ) \Bigr\} .
	\end{equation*}
Since $\hat{\Bz}$ belongs to this constraint set, 
$v_{h(|\cali{J}|)} \bigl(\hat z_{h(|\cali{J}|)}, 
\thetaJ_{h(|\cali{J}|)} \bigr) \leq \pi_{h(|\cali{J}|)}
(\bthetaJ)$.
Equality holds precisely when $\hat{\Bz}$ solves the 
final optimization problem, that is, when $\hat{\Bz} 
\in \BZ_{h(|\cali{J}|)}(\bthetaJ) = \BZ^*(\bthetaJ,h)$.

\smallskip

(c) From \autoref{C:local-j-properties}(c), the set 
$\BZ^*(\bthetaJ,h)$ is a singleton.  
Starting from the convex set $\BZ(\bthetaJ)$, 
strict concavity of $v_{h(1)}(\cdot,\thetaJ_{h(1)})$ 
implies that all maximizers at the first step of the 
recursion have the same trade position for the most 
powerful individual $h(1)$. 
The solution set is therefore obtained by fixing that 
individual's trade position and remains convex.  
Repeating this argument shows that $\BZ_{h(|\cali{J}|-1)}
(\bthetaJ)$ fixes the trade positions of the first 
$|\cali{J}|-1$ individuals in the hierarchy.  
Feasibility then fixes the weakest individual's trade 
position through $\sum_{i\in\cali{J}} z_i = \zero$.  
Hence this set is a singleton and coincides with 
$\BZ^*(\bthetaJ,h)$.  
By parts (a)--(b) of the proposition, every local jungle 
equilibrium belongs to this singleton, while 
\autoref{P:local-j-existence} shows that its unique
element is a local jungle equilibrium. 
This proves uniqueness at $(\bthetaJ,h)$.
\end{proof}


\section{General Existence of Occupational Equilibria}  
		\label{S:app-existence}

Fix a nonempty economy $\cali E \in \{\cali M, \cali J\}$ and
maintain \autoref{A:structure} as imposed on this economy.  
For the jungle, suppose that the conditions of 
\autoref{P:correspondence-j-continuity} hold. 
For the market, fix a commonly known continuous selection
$p \colon \BThetaM \to \R^L$ of the equilibrium price 
correspondence $P$.
We use the standard mixed-strategy formulation of continuous 
games \citep[Section~2]{Glicksberg:1952vp}.

Individual $i$'s strategy space is $\Delta(\ThetaE_i)$,  
the Borel probability measures on her occupational menu, 
endowed with the topology of weak convergence.
A strategy profile is then $\Bsig \, = \, (\sigma_i) 
\in \times_{i \in \cali E} \Delta(\ThetaE_i)$.  
Individuals draw their occupations independently according to 
these lotteries.  
The realized occupations determine their feasible trade sets 
and valuations. 
In the jungle, the hierarchy is drawn after occupations are 
realized and independently of the occupational lotteries, 
and the serial dictatorship then determines local jungle 
equilibrium trades.  
In the market, the fixed price selection $p$ assigns 
equilibrium prices to the realized occupational profile, 
and a corresponding local market equilibrium determines 
trades and monetary transfers.

Let $\nu_{\Bsig}$ denote the product probability law induced by
$\Bsig$ on $\BThetaE$.  
Individual $i$'s payoff in the occupational game with lotteries 
is then given by 
	\begin{equation*}   
		 \Pi_i(\Bsig)
		 \ \equiv \ \int_{\BThetaE} \bar\pi_i(\bthetaE) \,
			 d \nu_{\Bsig}(\bthetaE).  
	\end{equation*} 
Here $\bar\pi_i$ is the occupational payoff in the 
corresponding economy.  
Under the conditions imposed above, each $\bar\pi_i$ is 
continuous and bounded on the compact set $\BThetaE$, 
so the above integral is well defined, 
for $\cali E \in \{\cali M, \cali J\}$.

A (mixed) occupational equilibrium is a profile
$\Bsig^* \in \times_{i \in \cali E} \Delta(\ThetaE_i)$ 
such that
	\begin{equation}   \label{Eq:OE-mixed}
		\Pi_i(\sigma_i^*,\Bsig_{-i}^*)
		\ \geq \ \Pi_i(\sigma_i',\Bsig_{-i}^*),
			\qquad \text{for all $i \in \cali E$, all 
			$\sigma_i' \in \Delta (\ThetaE_i)$.}
	\end{equation}
Thus no individual gains by changing her occupational 
lottery while the others keep theirs fixed. 
A pure occupation $\thetaE_i$ is represented by the Dirac 
measure $\delta_{\thetaE_i}$, so pure
occupational equilibria are included in this formulation.

The mixed-strategy spaces are compact and convex, and 
$\Pi_i$ is continuous in the strategy profile and affine 
in $\sigma_i$. 
Consequently, the convexity of pure occupational best 
replies required for our pure-strategy existence results 
is unnecessary here, in either economy.

\begin{theorem}   \label{T:occup-existence}
Let $\cali E\in\{\cali M,\cali J\}$ be nonempty and suppose
that \autoref{A:structure} holds as imposed on $\cali E$.
In addition, assume that
\begin{enumerate}[leftmargin=1.5em,label=(\alph*)] 
		\setcounter{enumi}{0} \itemsep.25ex
	\item for $\cali E = \cali J$, the conditions of
    	\autoref{P:correspondence-j-continuity} hold;
	\item for $\cali E = \cali M$, occupational payoffs are 
		induced by a fixed, commonly known, continuous 
		selection $p \colon \BThetaM \to \R^L$ of the 
		equilibrium price correspondence $P$.
\end{enumerate}
Then the occupational game with strategy spaces
$\Delta(\ThetaE_i)$ and payoffs $\Pi_i$ admits an equilibrium
satisfying \autoref{Eq:OE-mixed}.
\end{theorem}

\begin{proof}
There are finitely many individuals, each menu $\ThetaE_i$ is a 
nonempty compact metric space, and each payoff $\bar\pi_i$ is 
continuous on $\BThetaE$. 
Glicksberg's existence theorem
\citep[Section~2]{Glicksberg:1952vp} therefore applies to this
occupational game and yields a profile of lotteries satisfying
\autoref{Eq:OE-mixed}.
\end{proof}


\end{document}